\documentclass[aps,prb, preprint, showpacs]{revtex4}
\usepackage{amsmath}
\usepackage{graphicx}

\input{tcilatex}
\begin{document}

\title{Density Functional Theory of Inhomogeneous Liquids: I. The
liquid-vapor interface in Lennard-Jones fluids.}
\author{James F. Lutsko}
\affiliation{Universite Libre de Bruxelles}

\begin{abstract}
A simple model is proposed for the direct correlation function (DCF) for
simple fluids consisting of a hard-core contribution, a simple parametrized
core correction and a mean-field tail. The model requires as input only the
free energy of the homogeneous fluid, obtained e.g. from thermodynamic
perturbation theory. Comparisons to the DCF obtained from simulation of a
Lennard-Jones fluid shows this to be a surprisingly good approximation for a
wide range of densities. The model is used to construct a density functional
theory for inhomogeneous fluids which is applied to the problem of
calculating the surface tension of the liquid-vapor interface. The numerical
values found are in good agreement with simulation.
\end{abstract}

\date{\today}
\maketitle

\section{Introduction}

The modern understanding of liquid-vapor interfaces begins with the seminal
paper of van der Waals in which he introduces what is now known as the
square-gradient approximation to the free energy of inhomogeneous systems
and the mean-field approximation\cite{VDW2,VDW1}. That work was developed
from a thermodynamic perspective which has been superseded by more
fundamental, statistical mechanical approaches\cite%
{EvansDFT,RowlinsonWidom,Lutsko_Physica_2006_2}. In the modern approach, it
is possible to give formally exact expressions for the free energy in terms
of structural quantities such as the pair-distribution function and the
direct correlation function. In practice, these expressions must be
approximated leading to a compromise between the two goals of simplicity and
accuracy. Given modern computational resources, simplicity is not an
overriding constraint when dealing with simple fluids governed by
spherically symmetric pair potentials but it quickly does become an issue
for more complex systems such as solids, fluids governed by anisotropic
potentials, extended molecules, etc. When confronted with these types of
complications, there is often little choice than to revert to the most
primitive mean field models and to hope that the qualitative picture
obtained is sufficient. The object of the present work is to present an
approach which is little more complex than the simplest mean field theory
and yet which gives quantitatively accurate results.

The original theory of van der Waals was based on two basic approximations.
First is what in modern language would be called a mean-field approximation
whereby the microscopic structure of the liquid is neglected and the only
spatial variation taken into consideration is a continuous variation of the
density. This means that the interaction energy for a system of molecules
interacting via a pair potential can be expressed in the form%
\begin{equation}
\int_{V}\int_{V}\rho \left( \mathbf{r}_{1}\right) \rho \left( \mathbf{r}%
_{2}\right) v\left( \mathbf{r}_{12}\right) d\mathbf{r}_{1}d\mathbf{r}_{2}
\label{1}
\end{equation}%
where $V$ is the volume of the system, $\rho \left( \mathbf{r}\right) $ is
the local density and $v\left( \mathbf{r}_{12}\right) $ is the pair
potential. The second is a gradient expansion about the center of mass. This
model was also discussed by Cahn and Hilliard\cite{CahnHilliard}. Today,
there are two more or less fundamental approaches to the description of
inhomogeneous liquids\cite{EvansDFT,RowlinsonWidom}. The oldest are the
integral equation methods where the object is to solve the Ornstein-Zernike
equation, which relates the pair-distribution function to the direct
correlation function, subject to an independent closure condition. This
approach is one of the most reliable methods for calculating the properties
of simple fluids. It is however intrinsically somewhat complex since the
fundamental objects being determined, e.g. the pair distribution function,
are two-body functions. The main alternative are density functional theories
which are somewhat simpler since the local density is a one-point function.
The utility of DFT\ for the description of liquid-vapor interfaces was first
demonstrated by the work of Ebner, Saam and Stroud\cite{Ebner-Saam}. They
used an approximate DFT which was, as they themselves later wrote\cite%
{Saam-Ebner}. somewhat ad hoc. The key quantity needed to evaluate the
theory was the direct correlation function (DCF)\ of the homogeneous fluid
which was obtained from the Percus-Yevick integral theory. However, for
interfacial calculations, the DCF was needed for all densities from that of
the liquid to that of the vapor which is problematic as the integral theory
does not possess solutions in the two-phase region. Ebner et al. were
therefore forced to interpolate the DCF's from the region where the integral
equation could be solved through the region without a solution. The
resulting values for the surface tension of a Lennard-Jones fluid were in
fact quite reasonable.

Since this early work, the DFT approach to interfacial problems has
developed primarily along two different lines. One is based on the
perturbative expression for the free energy in terms of a hard-sphere
contribution and a perturbative correction involving the potential and the
direct correlation function of hard spheres( see e.g. \cite{ZTang,Wadewitz}%
). This approach has the advantage that it reduces to the rather accurate
perturbative expression for the free energy for the homogeneous fluids and
therefore gives a good description of the phase diagram. Indeed, as
discussed by \cite{Wadewitz}, this is one of the main motivations for using
this model. The second approach is more in line with the work of Ebner et al
and is based on the direct correlation function (DCF)\ and the exact
relations between the DCF and the free energy. A good example is the recent
work of Tang and coworkers\cite%
{Tang-LJ-DFT-2003,Tang-LJ-DFT-2004,Tang-LJ-DFT-2005} who approximate the
DCF\ by that obtained from an approximate solution of the mean-spherical
approximation. (In the mean spherical approximation, the Ornstein-Zernike
equation is solved by assuming that there is an effective hard core inside
of which the pair distribution function vanishes and outside of which the
DCF is equal to $-\beta v\left( \mathbf{r}_{12}\right) $). To further
simplify, Tang approximates the Lennard-Jones interaction by a sum of Yukawa
potentials so that the resulting mean-spherical model can be solved
analytically both exactly\cite{Hoye77} and within the first order
approximation\cite{TANG-FMSA}. This is then used in an approximate DFT,
somewhat different from that of Ebner et al, to calculate a variety of
properties of inhomogeneous Lennard-Jones fluids including the surface
tension of the liquid-vapor interface\cite{Tang-LJ-DFT-2005} and the
structure of confined fluids\cite{Tang-LJ-DFT-2004},\cite{Tang-Yukawa-DFT}.
Since surface tension in particular is well-known to be very sensitive to
the range of the potential, the Yukawa's, which are short-ranged, must
eventually be replaced by the original potential in order to account for the
long-ranged contributions to the surface tension.

The perturbation-theory approach requires as input the pair-distribution
function of the reference system, usually hard-spheres for simple fluids.
Even when the reference fluid is hard-spheres, the pair-distribution
function is not an easy object to work with (see ref.\cite{Wadewitz}) and
the calculations would be even more difficult for more complex interactions.
On the other hand, the difficulty with the DCF-based theories is that there
have been few options for getting the DCF required in the theories:\ either
the mean-field approximation such as eq.(\ref{1}) above is used (which is
very crude) or the full machinery of liquid-state theory is used (which is
expensive). However, given that even when the latter approach is used, quite
ad hoc corrections, such as the interpolation of Ebner et al., are needed
suggests that the effort expended to try to produce as good a DCF as
possible is perhaps unnecessary. Instead, the work of Tang et al suggests
that the most important ingredient beyond the mean-field form is that the
underlying equation of state should be reasonably accurate. The present work
aims to test this intuition by studying a minimal extension of the
mean-field model for the DCF designed so as to reproduce a known equation of
state. Specifically, the model proposed here consists of a hard core,
described by the Fundamental Measure DFT\cite%
{Rosenfeld1,Rosenfeld2,tarazona_2000_1,WhiteBear}, a mean-field type tail
and a simple polynomial correction within the core with parameters chosen to
give the desired equation of state. The utility of the model lies in the
fact that the equation of state of the homogeneous bulk fluid, is much
easier to determine than are structural properties such as the DCF and the
pair distribution function and it is much less sensitive to the details of
the interaction potential. In this work, for application to the
Lennard-Jones fluid, first order thermodynamic perturbation theory is used.
This approach represents a relatively minimal extension of the mean field
model and as shown below, even a relatively crude model gives good results
for the surface tension of the liquid-vapor interface. It is in keeping with
the idea that the DCF should be a relatively simple object as illustrated,
e.g., by the contrast in hard-spheres as described by the Percus-Yevick
approximation where the pair-distribution function is a complicated function
with a lot of structure whereas the DCF is just a cubic polynomial within
the hard core. The same is true in the analytic solution of the
mean-spherical approximation for a sum of Yukawas\cite{Hoye77}.

In the next Section, the basic elements of Density Functional Theory are
reviewed. A plausible, but uncontrolled, approximation is introduced to give
a framework suitable for practical applications. It is shown that further
approximations give the functional used by Ebner et al.\cite{Ebner-Saam} as
well as that introduced by Tang\cite{Tang-LJ-DFT-2005} . Because it is still
often discussed in the literature, see e.g. refs. \cite{Cornelisse} and \cite%
{Duque-LJ-interface}, the square-gradient approximation is also described.
The third Section discusses the extended mean-field approximation for the
DCF. Different versions are described depending on how accurately the tail
of the DCF is modeled. These are compared to the simulation data of
Llano-Restrepo and Chapman\cite{DCF-Simulation} and it is demonstrated that
these models are in good agreement with the simulations. In the fourth
Section, the calculation of the surface tension for the liquid-vapor
interface of the Lennard-Jones fluid are presented. Aside from testing the
model for the DCF, the results from the different approximate DFTs are
compared and it is found that they are all in reasonable agreement with one
another and with the results from simulation. The paper ends with a
discussion of the results.

\section{Theory}

\subsection{Density Functional Theory formalism}

Density Functional Theory is based on the fact that there is a one-to-one
correspondence between applied external fields, $V_{ext}\left( \mathbf{r}%
\right) $, and the ensemble-averaged equilibrium density profile, $\rho
\left( \mathbf{r}\right) $. For a given external field, there is a
functional of the form 
\begin{equation}
\Omega \left[ n,V_{ext}\right] =F\left[ n\right] -\int \mu n\left( \mathbf{r}%
\right) d\mathbf{r}+\int V_{ext}\left( \mathbf{r}\right) n\left( \mathbf{r}%
\right) dr
\end{equation}%
such that $\Omega \left[ n,V_{ext}\right] $ is extremized by the equilibrium
density profile giving%
\begin{equation}
0=\left. \frac{\delta \Omega \left[ n,V_{ext}\right] }{\delta n\left( 
\mathbf{r}\right) }\right\vert _{n\left( \mathbf{r}\right) =\rho \left( 
\mathbf{r}\right) }=\left. \frac{\delta F\left[ n\right] }{\delta n\left( 
\mathbf{r}\right) }\right\vert _{n\left( \mathbf{r}\right) =\rho \left( 
\mathbf{r}\right) }-\mu +V_{ext}\left( \mathbf{r}\right) .  \label{EL1}
\end{equation}%
Throughout this Section, square brackets are used to denote a functional
dependence and round brackets to denote an ordinary function. In general,
the domain of the spatial integrals is unbounded with the effect of any
walls being explicitly accounted for by the external field. The presence of
a hard wall will, in this way, manifest itself by the equilibrium density
profile obtained from eq.(\ref{EL1}) giving zero density outside the wall.
However, for clarity, a volume $V$ will be explicitly indicated below with
the understanding that it simply connotes the region of non-zero density. The
functional $F$ is conveniently written as a sum of an ideal gas
contribution, 
\begin{equation}
\beta F_{id}\left[ n\right] =\int_{V\ }\left( n\left( \mathbf{r}\right) \log
\left( \Lambda ^{3}n\left( \mathbf{r}\right) \right) -n\left( \mathbf{r}%
\right) \right) d\mathbf{r,}
\end{equation}%
where $\Lambda $ is the thermal wavelength, and a remaining, excess
contribution $F_{ex}\left[ n\right] .$ In general, the latter is unknown,
but since it depends only on the density, and not explicitly on the field,
it can be expressed by expansion about a uniform state having constant
density $\overline{\rho }_{0}$ as%
\begin{eqnarray}
\frac{1}{V}\beta F_{ex}\left[ n\right] &=&\frac{1}{V}\beta F_{ex}\left( 
\overline{\rho }_{0}\right) +\beta \mu _{ex}\left( \overline{\rho }%
_{0}\right) \left( \overline{n}-\overline{\rho }_{0}\right) \\
&&-\frac{1}{V}\sum_{j=2}^{\infty }\frac{1}{j!}\int_{V}...\int_{V}\left(
n\left( \mathbf{r}_{1}\right) -\overline{\rho }_{0}\right) ...\left( n\left( 
\mathbf{r}_{j}\right) -\overline{\rho }_{0}\right) c_{j}\left( \mathbf{r}%
_{1},...,\mathbf{r}_{j};\overline{\rho }_{0}\right) d\mathbf{r}_{1}...d%
\mathbf{r}_{j}  \notag
\end{eqnarray}%
where $\mu _{ex}\left( \overline{\rho }_{0}\right) =\frac{\partial }{%
\partial \overline{\rho }_{0}}\frac{1}{V}F_{ex}\left( \overline{\rho }%
_{0}\right) $ and where $c_{j}\left( \mathbf{r}_{1},...,\mathbf{r}_{j};%
\overline{\rho }_{0}\right) $ is the j-body direct correlation function of a
uniform fluid. These are simply the functional derivatives of the $F_{ex}%
\left[ n\right] $ in the uniform limit, 
\begin{equation}
c_{j}\left( \mathbf{r}_{1},...,\mathbf{r}_{j};\overline{\rho }_{0}\right)
=\lim_{n\left( \mathbf{r}\right) \rightarrow \overline{\rho }%
_{0}}c_{j}\left( \mathbf{r}_{1},...,\mathbf{r}_{j};\left[ n\right] \right)
=-\lim_{n\left( \mathbf{r}\right) \rightarrow \overline{\rho }_{0}}\frac{%
\delta ^{j}\beta F_{ex}\left[ n,V_{ext}\right] }{\delta n\left( \mathbf{r}%
_{1}\right) ...\delta n\left( \mathbf{r}_{j}\right) }
\end{equation}%
and it can be shown that they correspond to the usual direct correlation
functions discussed in liquid state theory\cite{EvansDFT}. Thus, the free
energy functional of an arbitrary non-uniform system is expressed in terms
of the properties of a uniform fluid. This series can be resummed to give%
\begin{eqnarray}
\beta F_{ex}\left[ n\right] &=&\beta F_{ex}\left( \overline{\rho }%
_{0}\right) +V\beta \mu _{ex}\left( \overline{\rho }_{0}\right) \left( 
\overline{n}-\overline{\rho }_{0}\right) \\
&&-\int_{V}d\mathbf{r}_{1}\int_{V}d\mathbf{r}_{2}\int_{0}^{1}d\lambda
\int_{0}^{\lambda }d\lambda ^{\prime }\left( n\left( \mathbf{r}_{1}\right) -%
\overline{\rho }_{0}\right) \left( n\left( \mathbf{r}_{2}\right) -\overline{%
\rho }_{0}\right) c_{2}\left( \mathbf{r}_{1},\mathbf{r}_{2};\left[
n_{\lambda ^{\prime }}\right] \right)  \notag
\end{eqnarray}%
where $n_{\lambda }\left( \mathbf{r}\right) =\overline{\rho }_{0}+\lambda
\left( n\left( \mathbf{r}\right) -\overline{\rho }_{0}\right) $ the integral
depends on the two-body direct correlation function for an arbitrary density
profile. In particular, in the case of a uniform density $n(\mathbf{r})=%
\overline{n}$ the last term on the right gives the standard result for
homogeneous fluids,%
\begin{equation}
\frac{1}{V}\beta F_{ex}\left( \overline{n}\right) =\frac{1}{V}\beta
F_{ex}\left( \overline{\rho }_{0}\right) +\beta \mu _{ex}\left( \overline{%
\rho }_{0}\right) \left( \overline{n}-\overline{\rho }_{0}\right)
-\int_{V}\int_{0}^{1}\left( \overline{n}-\overline{\rho }_{0}\right)
^{2}c_{2}\left( r_{12};\overline{\rho }_{0}+\lambda \left( \overline{n}-%
\overline{\rho }_{0}\right) \right) \left( 1-\lambda \right) d\lambda d%
\mathbf{r}_{12},  \label{thermo}
\end{equation}%
which relates the DCF of the homogeneous system to the thermodynamics. (Note
that in writing this equation, the fact that the DCF of a simple fluid
depends only on the scalar $r_{12}=\left\vert \mathbf{r}_{1}-\mathbf{r}%
_{2}\right\vert $ has been explicitly indicated.) An important point in all
of these exact expressions is that the results for the free energy of the
system with density $n\left( \mathbf{r}\right) $ (or the liquid with density 
$\overline{n}$) are \emph{independent} of the choice of reference liquid
density $\overline{\rho }_{0}$. This is of course not true when
approximations are introduced but any approximation will involve implicitly
or explicitly a choice of reference density. In this Section, the reference
state has been explicitly indicated so as to make this clear.

While the n-body DCFs for a uniform system are in principle accessible, in
practice only quantities up to the two-body direct correlation function are
known with any confidence for arbitrary pair potentials using liquid state
theory such as the integral equations of Rogers and Young\cite{RY} and of
Hansen and Zerah\cite{Zarah}. Even then, the integral equations only possess
solutions for certain ranges of density and temperature. Furthermore, if the
goal is to develop a theory which can eventually be applied to more complex
systems involving asymmetric interactions or the solid phase, then this
approach is infeasible.

\subsection{Approximations to the exact theory}

In order to construct a more practical approach, it is first noted that the
only system for which good general approximations to the functional $F_{ex}%
\left[ n\right] $ exist is that of hard spheres. It is therefore useful to
consider the difference 
\begin{eqnarray}
\beta F_{ex}\left[ n\right] &=&\beta F_{ex}^{HS}\left[ n\right] +\beta
\Delta F_{ex}\left( \overline{\rho }_{0};d\right) +V\beta \Delta \mu
_{ex}\left( \overline{\rho }_{0};d\right) \left( \overline{n}-\overline{\rho 
}_{0}\right)  \label{exact} \\
&&-\sum_{j=2}^{\infty }\frac{1}{j!}\int_{V}...\int_{V}\left( n\left( \mathbf{%
r}_{1}\right) -\overline{\rho }_{0}\right) ...\left( n\left( \mathbf{r}%
_{j}\right) -\overline{\rho }_{0}\right) \Delta c_{j}\left( \mathbf{r}%
_{1},...,\mathbf{r}_{j};\overline{\rho }_{0}\right) d\mathbf{r}_{1}...d%
\mathbf{r}_{j}  \notag \\
&=&\beta F_{ex}^{HS}\left[ n\right] +\beta \Delta F_{ex}\left( \overline{%
\rho }_{0};d\right) +V\beta \Delta \mu _{ex}\left( \overline{\rho }%
_{0};d\right) \left( \overline{n}-\overline{\rho }_{0}\right)  \notag \\
&&-\int_{V}d\mathbf{r}_{1}\int_{V}d\mathbf{r}_{2}\int_{0}^{1}d\lambda
\int_{0}^{\lambda }d\lambda ^{\prime }\left( n\left( \mathbf{r}_{1}\right) -%
\overline{\rho }_{0}\right) \left( n\left( \mathbf{r}_{2}\right) -\overline{%
\rho }_{0}\right) \Delta c_{2}\left( \mathbf{r}_{1},\mathbf{r}_{2};\left[
n_{\lambda ^{\prime }}\right] \right)  \notag
\end{eqnarray}%
where $\Delta c_{j}\left( \mathbf{r}_{1},...,\mathbf{r}_{j};\overline{\rho }%
_{0}\right) =c_{j}\left( \mathbf{r}_{1},...,\mathbf{r}_{j};\overline{\rho }%
_{0}\right) -c_{j}^{HS}\left( \mathbf{r}_{1},...,\mathbf{r}_{j};\overline{%
\rho }_{0};d\right) $, etc. Then, the simplest nontrivial approximation is
to truncate the infinite series after the first term giving the theory
studied by Rosenfeld\cite{Rosenfeld-Pert_DFT},%
\begin{eqnarray}
\beta F_{ex}\left[ n\right] &\simeq &\beta F_{ex}^{HS}\left[ n\right] +\beta
\Delta F_{ex}\left( \overline{\rho }_{0};d\right) +V\beta \Delta \mu
_{ex}\left( \overline{\rho }_{0};d\right) \left( \overline{n}-\overline{\rho 
}_{0}\right) \\
&&-\frac{1}{2}\int_{V}\int_{V}\left( n\left( \mathbf{r}_{1}\right) -%
\overline{\rho }_{0}\right) \left( n\left( \mathbf{r}_{2}\right) -\overline{%
\rho }_{0}\right) \Delta c_{2}\left( r_{12};\overline{\rho }_{0}\right) d%
\mathbf{r}_{1}d\mathbf{r}_{2}.  \notag
\end{eqnarray}%
However, while this is suitable for some applications, it suffers from the
fact that the results depend on the choice of reference density, $\overline{%
\rho }_{0}$. In fact, in the uniform limit, it will not give the correct
free energy for the bulk fluid unless one demands that $\overline{\rho }_{0}=%
\overline{n}$. This also works for some inhomogeneous systems such as a
fluid in contact with a wall since there is a unique bulk limit far from the
wall\cite{Tang-LJ-DFT-2004}. However, in other problems, most notably that
of the planar liquid-vapor interface, there is no unique bulk limit and no
choice of reference density gives the correct bulk free energy in all bulk
regions\cite{Tang-LJ-DFT-2005}.

Perhaps the most natural approximation that is exact in the limit of a
homogeneous liquid is to replace the exact DCF for the inhomogeneous system
by the DCF of a homogeneous system evaluated at some position-dependent
density. There is considerable ambiguity in how to do this since the DCF is
a two-point function and the density is a one-point function. The only
formal requirements are that the DCF must be symmetric in its arguments and
it must reduce to the known result in the uniform limit. Perhaps the two
simplest approximations satisfying both requirements are%
\begin{equation}
\Delta c_{2}\left( \mathbf{r}_{1},\mathbf{r}_{2};\left[ n\right] \right)
\simeq \Delta c_{2}\left( r_{12};\frac{n\left( \mathbf{r}_{1}\right)
+n\left( \mathbf{r}_{2}\right) }{2}\right)  \label{choice1}
\end{equation}%
and%
\begin{equation}
\Delta c_{2}\left( \mathbf{r}_{1},\mathbf{r}_{2};\left[ n\right] \right)
\simeq \frac{1}{2}\left( \Delta c_{2}\left( r_{12};n\left( \mathbf{r}%
_{1}\right) \right) +\Delta c_{2}\left( r_{12};n\left( \mathbf{r}_{2}\right)
\right) \right) .  \label{choice2}
\end{equation}%
In the following, these will be referred to as the local DCF approximations
(LDCF-I and LDCF-II,\ respectively).

When the LDCF-I approximation is substituted into eq. (\ref{exact}), one
obtains after some rearrangement, 
\begin{eqnarray}
&&\beta F_{ex}\left[ n\right] =\beta F_{ex}^{HS}\left[ n\right] +\int d%
\mathbf{r}\;\beta \Delta f\left( n\left( \mathbf{r}\right) \right) \\
&&+\frac{1}{4}\int_{V}d\mathbf{r}_{1}\int_{V}d\mathbf{r}_{2}\left( n\left( 
\mathbf{r}_{1}\right) -n\left( \mathbf{r}_{2}\right) \right) ^{2}\Delta 
\overline{\overline{c}}_{2}\left( r_{12};\frac{n\left( \mathbf{r}_{1}\right)
+n\left( \mathbf{r}_{2}\right) }{2},\overline{\rho }_{0}\right)  \notag \\
&&-\frac{1}{2}\int_{V}d\mathbf{r}_{1}\int_{V}d\mathbf{r}_{2}\left[ 
\begin{array}{c}
\left( \frac{n\left( \mathbf{r}_{1}\right) +n\left( \mathbf{r}_{2}\right) }{2%
}-\overline{\rho }_{0}\right) ^{2}\Delta \overline{\overline{c}}_{2}\left(
r_{12};\frac{n\left( \mathbf{r}_{1}\right) +n\left( \mathbf{r}_{2}\right) }{2%
},\overline{\rho }_{0}\right) \\ 
-\left( n\left( \mathbf{r}_{1}\right) -\overline{\rho }_{0}\right)
^{2}\Delta \overline{\overline{c}}_{2}\left( r_{12};n\left( \mathbf{r}%
_{1}\right) ,\overline{\rho }_{0}\right)%
\end{array}%
\right]  \notag
\end{eqnarray}%
where $\Delta f\left( n\right) =\frac{1}{V}\beta F_{ex}\left( n\right) -%
\frac{1}{V}\beta F_{ex}^{HS}\left( n\right) $ is the difference in free
energy per unit volume of the homogeneous liquid at density $n$ and the
density of a homogeneous hard-sphere liquid at the same density and where 
\begin{equation}
\Delta \overline{\overline{c}}_{2}\left( r;n,\overline{\rho }_{0}\right)
\equiv 2\int_{0}^{1}\int_{0}^{\lambda }\Delta c_{2}\left( r;\overline{\rho }%
_{0}+\lambda ^{\prime }\left( n-\overline{\rho }_{0}\right) \right) d\lambda
^{\prime }d\lambda .
\end{equation}%
The LDCF-II\ gives a somewhat simpler expression,%
\begin{eqnarray}
\beta F_{ex}\left[ n\right] &=&\beta F_{ex}^{HS}\left[ n\right] +\int d%
\mathbf{r}\;\Delta f\left( n\left( \mathbf{r}\right) \right)  \label{dft} \\
&&+\frac{1}{2}\int_{V}\int_{V}\left( n\left( \mathbf{r}_{1}\right) -%
\overline{\rho }_{0}\right) \left( n\left( \mathbf{r}_{1}\right) -n\left( 
\mathbf{r}_{2}\right) \right) \Delta \overline{\overline{c}}_{2}\left(
r_{12};n\left( \mathbf{r}_{1}\right) ,\overline{\rho }_{0}\right) d\mathbf{r}%
_{2}d\mathbf{r}_{1}.  \notag
\end{eqnarray}%
This has the intuitively appealing form of the sum of a hard-sphere
contribution, a local free energy approximation and a nonlocal term that
explicitly depends, via the factor $\left( n\left( \mathbf{r}_{1}\right)
-n\left( \mathbf{r}_{2}\right) \right) $ , on the inhomogeneity of the
fluid. To make contact with earlier work, and anticipating the model for the
DCF discussed below, assume that the DCF can be written as the sum of a
short-ranged part, $\Delta \overline{\overline{c}}_{2}^{core}\left(
r_{12};n\left( \mathbf{r}_{1}\right) ,\overline{\rho }_{0}\right) $, and a
density-independent tail, $\Delta \overline{\overline{c}}_{2}^{tail}\left(
r_{12}\right) $. Then, the excess free energy can be written as%
\begin{eqnarray}
\beta F_{ex}\left[ n\right] &=&\beta F_{ex}^{HS}\left[ n\right] +\int d%
\mathbf{r}\;\Delta f\left( n\left( \mathbf{r}\right) \right) \\
&&+\frac{1}{4}\int_{V}\int_{V}\left( n\left( \mathbf{r}_{1}\right) -n\left( 
\mathbf{r}_{2}\right) \right) ^{2}\Delta c_{2}^{tail}\left( r_{12}\right) d%
\mathbf{r}_{2}d\mathbf{r}_{1}  \notag \\
&&+\frac{1}{2}\int_{V}\int_{V}\left( n\left( \mathbf{r}_{1}\right) -%
\overline{\rho }_{0}\right) \left( n\left( \mathbf{r}_{1}\right) -n\left( 
\mathbf{r}_{2}\right) \right) \Delta \overline{\overline{c}}%
_{2}^{core}\left( r_{12};n\left( \mathbf{r}_{1}\right) ,\overline{\rho }%
_{0}\right) d\mathbf{r}_{2}d\mathbf{r}_{1}  \notag
\end{eqnarray}%
Since the core contribution is assumed to be short ranged, is makes sense to
expand in terms of the difference $\left( n\left( \mathbf{r}_{1}\right)
-n\left( \mathbf{r}_{2}\right) \right) $ giving, see Appendix \ref{proof}
for details, 
\begin{eqnarray}
\beta F_{ex}\left[ n\right] &=&\beta F_{ex}^{HS}\left[ n\right] +\int d%
\mathbf{r}\;\Delta f\left( n\left( \mathbf{r}\right) \right)  \label{inter}
\\
&&+\frac{1}{4}\int_{V}\int_{V}\left( n\left( \mathbf{r}_{1}\right) -n\left( 
\mathbf{r}_{2}\right) \right) ^{2}\left[ 2\int_{0}^{1}\lambda \Delta
c_{2}\left( r_{12};\overline{\rho }_{0}+\lambda \left( \frac{n\left( \mathbf{%
r}_{1}\right) +n\left( \mathbf{r}_{2}\right) }{2}-\overline{\rho }%
_{0}\right) \right) d\lambda \right] d\mathbf{r}_{1}d\mathbf{r}_{2}  \notag
\\
&&+...  \notag
\end{eqnarray}%
where the neglected terms are integrals involving $\left( n\left( \mathbf{r}%
_{1}\right) -n\left( \mathbf{r}_{2}\right) \right) ^{n}$ for $n>2$.
Expanding $\Delta c_{2}$ about $\lambda =1$ , gives to leading order%
\begin{eqnarray}
\beta F_{ex}\left[ n\right] &=&\beta F_{ex}^{HS}\left[ n\right] +\int d%
\mathbf{r}\;\Delta f\left( n\left( \mathbf{r}\right) \right)  \label{Ebner}
\\
&&+\frac{1}{4}\int_{V}\int_{V}\left( n\left( \mathbf{r}_{1}\right) -n\left( 
\mathbf{r}_{2}\right) \right) ^{2}\Delta c_{2}\left( r_{12};\frac{n\left( 
\mathbf{r}_{1}\right) +n\left( \mathbf{r}_{2}\right) }{2}\right) d\mathbf{r}%
_{1}d\mathbf{r}_{2}  \notag \\
&&+...  \notag
\end{eqnarray}%
which resembles the well-known theory of Ebner et al.\cite%
{Ebner-Saam,Saam-Ebner}. (In the original work, the hard-sphere contribution
was not treated separately. However, because the approximation scheme used
here for the excess part is the same, this will be referred to as the ESS
theory after the authors of the first paper.) If instead one expands $\Delta
c_{2}$ about $\lambda =0$ and keeps only the leading term, the result is%
\begin{eqnarray}
\beta F_{ex}\left[ n\right] &=&\beta F_{ex}^{HS}\left[ n\right] +\int d%
\mathbf{r}\;\beta \Delta f\left( n\left( \mathbf{r}\right) \right) \\
&&+\frac{1}{4}\int_{V}\int_{V}\left( n\left( \mathbf{r}_{1}\right) -n\left( 
\mathbf{r}_{2}\right) \right) ^{2}\Delta c_{2}\left( r_{12};\overline{\rho }%
_{0}\right) d\mathbf{r}_{1}d\mathbf{r}_{2}  \notag \\
&&+...  \notag
\end{eqnarray}%
which is the recent theory of Tang\cite{Tang-LJ-DFT-2005}.

Finally, having outlined the approximations that will be used in the
applications below, it is worth noting some formal differences between them.
The LDCF approximations involve a minimal conceptual element, eq.(\ref%
{choice1}) or eq.(\ref{choice2}), but as anticipated above and as shown
explicitly below, the resulting free energy is no longer independent of the
chosen reference state. For the liquid-vapor interface, it will turn out
that the dependence on $\overline{\rho }_{0}$ is quite weak except that if
it is chosen too large, no stable profile is found. For at least this
application, these theories are practically, if not formally, unique. The
same is true of the Tang theory although the dependence on the reference
state is found to be somewhat stronger than for the LDCF theories. The ESS
theory is independent of the reference state. However, hidden in its
derivation is an expansion of the density integrals in eq.(\ref{inter})
about an arbitrarily chosen point ($\lambda =1$). Perhaps it could be argued
that the expansion about $\lambda =1$ is justified by the fact that it makes
the theory independent of the reference state, but whether or not this is
convincing seems to be a matter of taste.

\subsection{Square-gradient approximation}

Another approach to the description of inhomogeneous systems is the
square-gradient approximation\cite%
{VDW2,VDW1,RowlinsonWidom,Lutsko_Physica_2006_2}. If the density varies
sufficiently slowly, it is possible to expand the density-dependence of the
exact free energy functional so as to obtain%
\begin{equation}
\beta F_{ex}\left[ n\right] =\int_{V}\;\left[ \beta f\left( n\left( \mathbf{r%
}\right) \right) +\frac{1}{2}g\left( n\left( \mathbf{r}\right) \right)
\left( \mathbf{\nabla }n\left( \mathbf{r}\right) \right) ^{2}+...\right] d%
\mathbf{r}
\end{equation}%
where the neglected terms involve higher order derivatives. The coefficient
of the gradient term is 
\begin{equation}
g\left( n\right) =\frac{1}{6}\int_{V}r^{2}c_{2}\left( r;n\right) d\mathbf{r}
\end{equation}%
showing that the square-gradient approximation is another way to use
information about the uniform system to construct a description of
nonuniform systems. The density profile is determined by the Euler-Lagrange
equation%
\begin{equation}
\mathbf{\nabla }\cdot \left( g\left( n\left( \mathbf{r}\right) \right) 
\mathbf{\nabla }n\left( \mathbf{r}\right) \right) -\frac{\partial g\left(
n\left( \mathbf{r}\right) \right) }{\partial n\left( \mathbf{r}\right) }%
\frac{1}{2}\left( \mathbf{\nabla }n\left( \mathbf{r}\right) \right) ^{2}-%
\frac{\partial }{\partial n\left( \mathbf{r}\right) }\left( \beta f\left(
n\left( \mathbf{r}\right) \right) -\mu n\left( \mathbf{r}\right) \right) =0.
\end{equation}%
In the original theory of van der Waals, the dependence of $g$ on the
density was neglected and the resulting constant value of $g$ is known as
the influence parameter.

\section{The extended mean field mode for the DCF}

\subsection{Formulation of the model}

In order to apply any of the approximate DFT's discussed above, it is
necessary to know the DCF. The only general approach to determine DCF's up
to liquid densities is via integral equation theory. However, this can be
computationally expensive for complex systems and also suffers from the fact
that solutions often do not exist for some combinations of density and
temperature. For a homogeneous system, this is not a problem if the DFT
being used involves only the local density (as in the approximations of ESS
and of Tang). But for application to liquid-vapor interfaces, the whole
range of densities from liquid to vapor, necessarily including densities in
the two-phase region. As discussed in the Introduction, the ad hoc nature of
the solutions to this problem suggest that the DCF need not be so precisely
determined for the purposes of DFT. Thus, the goal here is to put together
as simple a model as possible that preserves certain basic exact properties
of the DCF. The basic structure of the models considered is 
\begin{equation}
c\left( r;n\right) =c_{HS}\left( r;n,d\right) +\Theta \left( d-r\right)
\left( a_{0}\left( n,T\right) +a_{1}\left( n,T\right) \frac{r}{d}\right)
+c^{tail}\left( r;n,d\right) ,
\end{equation}%
where the first term is the DCF for a hard-sphere system with hard-sphere
diameter $d$, the second term is a correction to the hard-sphere part in the
core region and the third part is the "tail"\ of the distribution. The
hard-sphere DCF will be chosen to be consistent with the hard-sphere theory
used in the DFT. For simple fluids, this will mean either the Percus-Yevick
DCF or the one associated with the "White-Bear"\ FMT\cite{WhiteBear}. Both
of these are only nonzero for $r<d$ and in the core region they are cubic
polynomials in $r$. The core-correction is shown as a linear polynomial,
although there is no reason that some other form could not be used. The
coefficients will be determined by demanding that the DCF agree with a known
equation of state, via eq.(\ref{thermo}), and by the requirement that the
DCF be continuous, as it is expected to be for any continuous potential.
This gives%
\begin{equation}
4\pi d^{3}\left( \frac{1}{3}a_{0}\left( n,T\right) +\frac{1}{4}a_{1}\left(
n,T\right) \right) =\frac{\partial ^{2}f_{ex}^{HS}\left( n\right) }{\partial
n^{2}}-\frac{\partial ^{2}f_{ex}\left( n\right) }{\partial n^{2}}-4\pi
\int_{0}^{\infty }c^{tail}\left( r;n,d\right) r^{2}dr
\end{equation}%
and%
\begin{equation*}
c_{HS}\left( d_{-},n,d\right) +a_{0}\left( n,T\right) +a_{1}\left(
n,T\right) +c^{tail}\left( d_{-};n,d\right) =c^{tail}\left( d_{+};n,d\right)
\end{equation*}%
where $d_{\pm }$ refers to the limit $r\rightarrow d$ from above or below.
Note that the core correction could include higher order terms with
continuity of the first and higher order derivatives being used to determine
the coefficients but only the minimal model involving the linear correction
will be studied here. It seems natural to refer to this combination of
hard-core +\ core correction +\ tail as an "extended mean field model".
While any reasonable value for the hard-sphere diameter could be used, the
calculations presented below are based on the Barker-Henderson formula\cite%
{BarkerHend,HansenMcdonald},%
\begin{equation}
d=\int_{0}^{r_{0}}\left( 1-\exp \left( -\beta v\left( r\right) \right)
\right) dr,  \label{BH-HSD}
\end{equation}%
where $r_{0}$ is the point at which the potential equals zero.

In order to fix the form of the tail function, there is one useful piece of
information that can be considered:\ namely, that at zero density, the DCF
is known to be equal to the negative of the Mayer function, 
\begin{equation}
c\left( r;\overline{n}=0\right) =\exp \left( -\beta v\left( r\right) \right)
-1.  \label{meyer}
\end{equation}%
This suggests the low-density model in which the tail function is simply the
difference between the Mayer function for the interaction potential and that
for hard-spheres giving%
\begin{equation}
c^{tail}\left( r;n,d\right) =\exp \left( -\beta v\left( r\right) \right)
-1+\Theta \left( d-r\right)
\end{equation}%
In this case, if the equation of state has the property that it gives the
exact second virial coefficient in the low density limit, then the model DCF
will reduce to the exact result in that limit. Unfortunately, while some
versions of thermodynamic perturbation theory do possess this property (e.g.
that of Paricaud\cite{Paricaud_Correct_B2}), some of the most well-known
theories, such as those of Barker and Henderson\ (BH)\cite%
{BarkerHend,HansenMcdonald}and that of Weeks, Chandler and Andersen (WCA) 
\cite{WCA1,WCA2,WCA3,HansenMcdonald}, do not and so the low-density model
will not, in this case, give the correct behavior.

If the exact low-density limit cannot be enforced because of inadequacies in
the equation of state, then it may be reasonable to forgo the complexity of
the Mayer function. In thermodynamic perturbation theory, the potential is
typically written as a sum of a short ranged-repulsion, $v_{0}\left(
r\right) $ and a long ranged attraction, $w\left( r\right) $, and it is
assumed that the repulsive part can be well approximated by the hard-sphere
potential. Applying this to the low-density tail gives%
\begin{eqnarray}
c^{tail}\left( r;n,d\right) &=&\exp \left( -\beta v_{0}\left( r\right)
\right) \exp \left( -\beta w\left( r\right) \right) -1+\Theta \left(
d-r\right) \\
&\simeq &\Theta \left( r-d\right) \exp \left( -\beta w\left( r\right)
\right) -1+\Theta \left( d-r\right)  \notag \\
&\simeq &-\Theta \left( r-d\right) \beta w\left( r\right)  \notag
\end{eqnarray}%
where the last line is a good approximation at high temperatures. In fact,
in perturbation theory, it is often found that this type of approximation is
also accurate at high densities, regardless of the temperature, so that this
mean-field approximation is frequently more useful than these arguments
would suggest\cite{Stell_pt_exact}. In the following, this approximation
will be used with the simplest choice of the long-ranged part of the
potential, namely $w(r)=v(r)$.

Finally, we discuss an interpolation between the low-density model and the
extended mean-field model. The former is exact at low density, provided the
equation of state gives the correct second virial coefficient. However, at
higher densities, it is often better to assume the mean-field tail as
opposed to the Mayer function tail (this is in part the rational behind the
Mean Spherical Approximation). The same type of thing occurs in
thermodynamic perturbation theory where a "resummed" perturbation theory is
sometimes used that interpolates between these two forms\cite%
{BHMacroscopicCompressibility,BarkerHend_WhatIsLiquid,Paricaud_Correct_B2}.
The equivalent idea here would be to represent the tail of the DCF as 
\begin{equation}
c^{tail}\left( r;n,d\right) =\exp \left( -\beta v_{0}\left( r\right) \right)
\left( 1+\kappa _{HS}^{-1}\left( \overline{\rho }\right) \left( \exp \left(
-\kappa _{HS}\left( \overline{\rho }\right) \beta w\left( r\right) \right)
-1\right) \right) -1+\Theta \left( d-r\right)
\end{equation}%
where the potential has again been separated into a short-ranged repulsion, $%
v_{0}\left( r\right) $, and a long ranged attraction, $w\left( r\right) $,
as in perturbation theory( see, e.g. ref. \cite{HansenMcdonald} ). The
function $\kappa _{HS}\left( \overline{\rho }\right) =\left. \frac{\partial 
\overline{\rho }}{\partial \beta P}\right\vert _{T}$ is the reduced
compressibility of a hard-sphere system at density $\overline{\rho }$ . At $%
\overline{\rho }=0$, the compressibility is one and this is identical to the
low-density approximation giving the negative of the Mayer function. At high
density, the compressibility becomes very small and this becomes very
similar to the mean-field tail. The use of the hard-sphere compressibility
to control the switching occurs naturally in perturbation theory\ (see refs. 
\cite%
{BHMacroscopicCompressibility,BarkerHend_WhatIsLiquid,Paricaud_Correct_B2})
although here, it is simply adopted as a convenient model. The model with
this form of the tail will be referred to as the hybrid model. In the
calculations discussed below, it is implemented using the usual WCA
separation of the potential\cite{WCA1,WCA2,WCA3,HansenMcdonald}.

\subsection{Comparison to simulation}

In this Section, the extended mean field model for the direct correlation
function is illustrated by comparison to data from molecular dynamics
simulations for a Lennard-Jones potential,%
\begin{equation}
v\left( r\right) =4\varepsilon \left( \left( \frac{\sigma }{r}\right)
^{12}-\left( \frac{\sigma }{r}\right) ^{6}\right) .
\end{equation}%
The equation of state of the bulk fluid was calculated using first-order
thermodynamic perturbation theory using both the BH and WCA theories and the
resulting phase diagrams are shown in Fig. 1. The Barker-Henderson theory
gives somewhat low liquid densities and higher vapor densities with a lower
critical point than does the WCA theory. Neithr theory is very accurate near
the critical point where renormalization effects are expected to be
important.

Figure 2 shows a comparison between the exact DCF at zero density and a
reduced temperature of $T^{\ast }\equiv k_{B}T/\varepsilon =1.5$ and that of
the model. In Fig, 2a, the model is evaluated using the exact second virial
coefficient so that the low-density and hybrid tails reproduce the exact
result. The error found in using the mean-field tail is due to compensation
in the core for the errors made outside the core in the integral of the DCF.
Note that the thermodynamic constraint depends on the spatial integral of
the DCF times $r^{2}$ which explains the relatively large deviations
required inside the core. Figure 2b shows the models as evaluated using the
Barker-Henderson theory. Since the low-density free energy is incorrect in
this theory, i.e. the second virial coefficient is wrong, the low-density
and hybrid models now include spurious core corrections whereas the core
correction for the mean-field tail is actually smaller. This is because the
mean-field tail is very close to that which is used in the Barker-Henderson
perturbation theory giving a case of compensating errors.

Figures (3)-(5)\ show the DCF for different densities for $T^{\ast }=1.5$ and Fig. (6)
shows the DCF for $T^{\ast }=0.72$ and $\rho \sigma ^{3}=0.72$ . While this
simple model cannot be expected to be perfect, the figures show that it
represents a good first approximation to the actual DCF. For the higher
temperature, the low-density tail appears slightly better at the lowest
density. At the lower temperature, the difference is greater. This is
because the peak in the DFC is due to the minimum in the potential and the
low density tail, due to the exponentiation of the potential, gives a higher
peak than both the mean field tail and the data. In order to give the same
integral, the core correction is therefore forced to be more negative.
Altogether, it would appear that all of the tail approximations are
reasonable. The low-density tail is better at low densities, the mean-field
tail is better at high density and the hybrid model is overall the most
accurate. Nevertheless, it would appear that the additional analytic
complexity of the low-density and hybrid tails are only justified if the
equation of state is exact at low density and if there is particular
interest in reproducing the exact low-density DCF.

For the highest densities, see Figs.(5) and (6), the hard-sphere
contribution to the DCF is also shown so as to highlight the role of
the core correction. For $T^{\ast}=1.5$, the
core correction gives a modest improvement over the
hard-sphere DCF over most of the core region. The errors are largest
at $r=0$ which is probably the least important region. On the other hand,
for $T^{\ast}=0.72$,
the core correction gives a clear improvement over
the hard-sphere DCF throughout the entire core region.

\bigskip

\section{Application to the planar interface}

\subsection{Reduction to a one-dimensional problem}

In this Section, the model DCF is used to evaluate the density functional
theories discussed in Section II for the case of a planar liquid-vapor
interface. The discussion here will focus on the extended mean-field
approximation for the DCF with the mean-field tail. Calculations with the
more complex models require more numerical analysis and confirm the
conclusions of Section III that the additional complexity has little effect
on the quantitative results.

The density is assumed to vary in only one direction, say the $z$
-direction, and to be constant in all other directions. For a fixed
temperature below the critical point, there is a unique value of the liquid
and vapor densities, $\overline{n}_{l}$ and $\overline{n}_{v}$ , such that
the two phases have the same chemical potential and pressure and can
therefore coexist. Within the LDCF model, the excess free energy per unit
area, i.e. the surface tension, can be written as%
\begin{eqnarray}
\gamma \equiv \frac{1}{A}\left( \Omega \left[ n\right] -\Omega \left( 
\overline{n}_{l}\right) \right) &=&\beta F_{ex}^{HS}\left[ n\right]
+\int_{-\infty }^{\infty }\left( \Delta f\left( n\left( z\right) \right)
-\mu n\left( z\right) -\left( f\left( n_{l}\right) -\mu n_{l}\right) \right)
dz  \label{tau} \\
&&+\frac{1}{4}\int_{-\infty }^{\infty }\int_{-\infty }^{\infty }\left(
n\left( z_{1}\right) -\overline{\rho }_{0}\right) \left( n\left(
z_{1}\right) -n\left( z_{2}\right) \right) \widetilde{c}_{2}^{core}\left(
z_{12};n\left( z_{1}\right) ,\overline{\rho }_{0}\right) dz_{2}dz_{1}  \notag
\\
&&+\frac{1}{4}\int_{-\infty }^{\infty }\int_{-\infty }^{\infty }\left(
n\left( z_{1}\right) -n\left( z_{2}\right) \right) ^{2}\beta \widetilde{v}%
\left( z_{12}\right) dz_{2}dz_{1}  \notag
\end{eqnarray}%
where 
\begin{equation}
\widetilde{v}\left( z\right) \equiv \int_{-\infty }^{\infty }\int_{-\infty
}^{\infty }v\left( r\right) \Theta \left( r-d\right) dxdy
\end{equation}%
and%
\begin{equation}
\widetilde{c}_{2}^{core}\left( z;n,\overline{\rho }_{0}\right) \equiv
\int_{-\infty }^{\infty }\int_{-\infty }^{\infty }\left( \overline{\overline{%
a}}_{0}\left( n,\overline{\rho }_{0}\right) +\overline{\overline{a}}%
_{1}\left( n,\overline{\rho }_{0}\right) \frac{r}{d}\right) \Theta \left(
d-r\right) dxdy.
\end{equation}%
are the planar averages of the potential and the core correction to the DCF. The constants in
the core term are related to those in the extended mean-field model by%
\begin{equation}
\overline{\overline{a}}_{i}\left( n,\overline{\rho }_{0}\right)
=2\int_{0}^{1}d\lambda \int_{0}^{\lambda }d\lambda ^{\prime }\;a_{i}\left( 
\overline{\rho }_{0}+\lambda ^{\prime }\left( n-\overline{\rho }_{0}\right)
\right) .
\end{equation}%
Note that the equivalent of the model of ESS is obtained by the substitution 
$\overline{\overline{a}}_{i}\left( n,\overline{\rho }_{0}\right) \rightarrow 
\frac{1}{2}a_{i}\left( n\right) $ whereas that of Tang results from $%
\overline{\overline{a}}_{i}\left( n,\overline{\rho }_{0}\right) \rightarrow
a_{i}\left( \overline{\rho }_{0}\right) $ together with the specific choice $%
\overline{\rho }_{0}=\frac{1}{2}\left( \overline{n}_{l}+\overline{n}%
_{v}\right) $. The equilibrium density profile is found by minimizing the
free energy with respect to the density profile subject to the boundary
conditions 
\begin{eqnarray}
\lim_{z\rightarrow -\infty }n\left( z\right) &=&n_{l}\left( \mu ,T\right) \\
\lim_{z\rightarrow \infty }n\left( z\right) &=&n_{v}\left( \mu ,T\right) 
\notag \\
\lim_{z\rightarrow \pm \infty }\frac{d}{dz}n\left( z\right) &=&0.  \notag
\end{eqnarray}

\subsection{Implementation}

The numerical work begins with the calculation of the Barker-Henderson
hard-sphere diameter, $d(T)$, from eq.(\ref{BH-HSD}). The coefficients $%
a_{i}\left( n\right) $ are evaluated for 100 points in the density range $%
0\leq n\sigma ^{3}\leq 1$ using either the BH or WCA first order
perturbation theories. They are then interpolated using cubic splines which
permit easy calculation of the coefficients $\overline{\overline{a}}%
_{i}\left( n,\overline{\rho }_{0}\right) $ for a given value of $\overline{%
\rho }_{0}$. (In this regard, the difference in computational complexity
between the LDCF theory and the approximations of ESS and Tang is minimal.)
The calculation of the free energy is discretized by introducing a lattice
of points on the interval $[-L,L]$ via $z_{i}=-L+i\delta $ for $i=0$ to $%
N=2L/\delta $. The limits are chosen sufficiently large that it may be
assumed that $n\left( L\right) =n_{v}$ and $n\left( -L\right) =n_{l}$ and
the goal is to find the profile in the form of the points $n_{i}=n\left(
z_{i}\right) $ which minimize the free energy functional. (Although
care must be taken to include the contributions of the regions outside
this range, which can be done analytically.) The minimimization of the
free energy was performed using the Broyden-Fletcher-Goldfarb-Shanno quasi-Newton
method as implemented in the GNU\ Scientific Library\cite{GSL}.  Except where otherwise noted, all results reported
here are based on a lattice of $20$ points per hard-sphere diameter, $\delta
=d/20$.

\subsection{Results}

Figure 7 shows the reduced surface tension, $\gamma^{\ast}=\gamma\sigma^{2}/%
\varepsilon$ as a function of temperature obtained using the LDC-II with
both the BH and WCA first order perturbation theories for the equations of
state, as well as the simulation results of Mecke, Winkelmann and Fischer%
\cite{Mecke-LJ_Interface} of Duque, Pamies and Vega\cite{Duque-LJ-interface}
and of Potoff and Panagiotopoulos\cite{Potoff_LJ_Interface}. The surface
tensions calculated are quite sensitive to the densities of the coexisting
phases so, as might be guessed from the phase diagrams, the BH results are
more accurate at higher temperatures, near the critical point, whereas at
the lowest temperatures, the results using both perturbation theories are
comparable. Figure 8 shows the density profiles obtained for $T^{\ast }=0.7$
calculated using all of the various DFTs discussed above. The SGA\ gives a
much broader profile than the other DFTs. The LDCF and Tang models give
smooth profiles while the ESS shows some small oscillations at the
transition to high density, but the profiles for all four models are
extremely similar. They all show a very rapid increase in density moving
from the vapor into the liquid, followed by a slower, more rounded profile
as the density approaches that of the liquid.

Figure 9 shows a comparison of several different versions of the DFT
including the LDCF-I, eq.(\ref{choice1}), the LDCF-II,eq.(\ref{choice2}),
and the approximate theories of ESS type, Tang and the Square-Gradient
Approximation (SGA). With one exception, all of the DFTs except the SGA give
almost identical results. The SGA gives a considerably higher surface
tension, a fact well-known in the literature\cite{Cornelisse}. The exception
is that in the present calculations, the ESS theory is unstable at $T^{\ast
}=0.6$ and no solution was found.

The exact free energy is independent of the path in density space used to
calculate it. This property is shared in the ESS-type theory but the other
approximate DFTs all contain an explicit dependence on the reference
density. Figure 10 shows the variation of the surface tension as a function
of the reference density for $T^{\ast }=0.7$ . Note that the LDCF
approximations as well as that of Tang do not give stable solutions if the
reference density is chosen too large. Within the region that solutions
exist, the resulting free energies are in fact only weakly sensitive to the
value of the reference density, with the Tang theory showing the largest
variation. The LDCF-I theory is the most robust in the sense of possessing
the widest range of possible values of reference density.

One interesting distinction between the theories is the range of
temperatures for which they give stable solutions. The liquid-vapor
interface only exists for temperatures below the critical point (around $%
T^{\ast }=1.3$ in simulation and the BH\ theory). As the temperature is
lowered, the system eventually reaches the triple point, at about $T^{\ast
}=0.7$, below which the thermodynamically stable phases are the vapor and
the solid. However, the liquid can still exist as a metastable phase and
indeed, the usual perturbation theories continue to work for temperatures
much lower than the triple point. It is therefore of interest to check the
behavior of the DFT's for lower temperatures where a metastable liquid-vapor
interface is possible. In fact, both of the LDCF theories continue to give
sensible results for temperatures as low as $T^{\ast }=0.25$ as does the
model of Tang. On the other hand, attempts to solve the ESS-theory below the
triple point are problematic. Calculations performed at $T^{\ast }=0.6$ with
the ESS\ are stable but the profile shows significant oscillations. However,
when the density of lattice points is doubled, it is no longer possible to
find a stable solution and this is true at lower temperatures, even using
the original lattice spacing. For comparison, halving the lattice spacing
has no qualitative affect on the LDCF calculations, even at the lowest
temperature and the surface tension changes by less than $0.2\%$.

\section{Conclusions}

The extended mean field model for the DCF, consisting of a hard-core
contribution, a mean-field tail and a linear core correction, has been shown
to be a reasonable approximation to the DCF for a Lennard-Jones fluid. The
model retains much of the simplicity of the mean-field model but is
constructed to give a faithful representation of the thermodynamics of the
homogeneous system. The approach used here differs somewhat from that
commonly found in the literature wherein the goal is to make an ab initio
ansatz for DCF, or more generally, for the DFT, which is subsequently tested
by comparing its prediction of the properties of the homogeneous system to
simulation. Given that good, computationally efficient means exist - and
have long existed - for calculating such properties, there is no real reason
to try to construct a DFT at this level. Instead, the philosophy used here
is to view DFT as a to0l which is primarily useful for studying more complex
inhomogeneous systems and which, as such, is legitimately constructed
assuming a priori knowledge of the homogeneous system.

In this paper, the model DCF was used in conjunction with several
approximate DFTs to study the liquid-vapor interface of a Lennard-Jones
fluid. It was found that aside from the well-known exception of the SGA, all
of the DFTs gave very similar results for both the surface tension and the
density profile. The calculated surface tensions were also found to agree
well with the results from simulations.

It was noted at several points that the different approximate DFT's were
sometimes unstable in the sense that no smooth density profile could be
obtained. Fundamentally, this is due to the fact that the hard-sphere part
of the free energy, described by FMT, involves smoothed-densities  while the attractive part of the free energy involves
the density evaluated at a point. The instabilities arise because a very
localized spike in the density can increase the size of the attractive part
of the free energy whereas, because of the smoothing, it has little effect
on the hard-sphere contribution. This defect of the LDCF\ theories will be
explored further in a future publication. Further applications of this work
will be to the study of different potential models and geometries,
particularly the case of anisotropic potentials.

\begin{acknowledgments}
I am grateful to Marc Baus for several useful comments on an early draft of
this paper. This work was supported in part by the European Space Agency
under contract number ESA AO-2004-070.
\end{acknowledgments}

\bigskip \bigskip \appendix{}

\section{\protect\bigskip Proof of eq.(\ref{inter})}

\label{proof}

To prove eq.(\ref{inter}), the two-body term is written as%
\begin{eqnarray}
&&-\frac{1}{2}\int_{V}\int_{V}\left( n\left( \mathbf{r}_{1}\right) -%
\overline{\rho }_{0}\right) \left( n\left( \mathbf{r}_{2}\right) -n\left( 
\mathbf{r}_{1}\right) \right) \Delta \overline{\overline{c}}_{2}\left(
r_{12};n\left( \mathbf{r}_{1}\right) ,\overline{\rho }_{0}\right) d\mathbf{r}%
_{1}d\mathbf{r}_{2} \\
&=&\frac{1}{4}\int_{V}\int_{V}\left( n\left( \mathbf{r}_{2}\right) -n\left( 
\mathbf{r}_{1}\right) \right) ^{2}\Delta \overline{\overline{c}}_{2}\left(
r_{12};n\left( \mathbf{r}_{1}\right) ,\overline{\rho }_{0}\right) d\mathbf{r}%
_{1}d\mathbf{r}_{2}  \notag \\
&&-\frac{1}{2}\int_{V}\int_{V}\left( \frac{n\left( \mathbf{r}_{1}\right)
+n\left( \mathbf{r}_{2}\right) }{2}-\overline{\rho }_{0}\right) \left(
n\left( \mathbf{r}_{2}\right) -n\left( \mathbf{r}_{1}\right) \right) \Delta 
\overline{\overline{c}}_{2}\left( r_{12};\frac{n\left( \mathbf{r}_{1}\right)
+n\left( \mathbf{r}_{2}\right) }{2}+\frac{n\left( \mathbf{r}_{1}\right)
-n\left( \mathbf{r}_{2}\right) }{2},\overline{\rho }_{0}\right) d\mathbf{r}%
_{1}d\mathbf{r}_{2}  \notag
\end{eqnarray}%
Expanding in the difference in densities gives%
\begin{eqnarray}
&&\Delta \overline{\overline{c}}_{2}\left( r_{12};\frac{n\left( \mathbf{r}%
_{1}\right) +n\left( \mathbf{r}_{2}\right) }{2}+\frac{n\left( \mathbf{r}%
_{1}\right) -n\left( \mathbf{r}_{2}\right) }{2},\overline{\rho }_{0}\right) 
\\
&=&\int_{0}^{1}dx\left( 1-x\right) \Delta c_{2}\left( r_{12};\overline{\rho }%
_{0}+x\left( \frac{n\left( \mathbf{r}_{1}\right) +n\left( \mathbf{r}%
_{2}\right) }{2}+\frac{n\left( \mathbf{r}_{1}\right) -n\left( \mathbf{r}%
_{2}\right) }{2}-\overline{\rho }_{0}\right) \right)   \notag \\
&=&\Delta \overline{\overline{c}}_{2}\left( r_{12};\frac{n\left( \mathbf{r}%
_{1}\right) +n\left( \mathbf{r}_{2}\right) }{2},\overline{\rho }_{0}\right) 
\notag \\
&&+\left( \frac{n\left( \mathbf{r}_{1}\right) -n\left( \mathbf{r}_{2}\right) 
}{2}\right) \int_{0}^{1}dx\left( 1-x\right) x\left. \frac{\partial \Delta 
\overline{\overline{c}}_{2}\left( r_{12};n\right) }{\partial n}\right\vert _{%
\overline{\rho }_{0}+x\left( \frac{n\left( \mathbf{r}_{1}\right) +n\left( 
\mathbf{r}_{2}\right) }{2}-\overline{\rho }_{0}\right) }  \notag \\
&&+O\left( \frac{n\left( \mathbf{r}_{1}\right) -n\left( \mathbf{r}%
_{2}\right) }{2}\right) ^{2}  \notag
\end{eqnarray}%
The second term is 
\begin{eqnarray}
&&\int_{0}^{1}dx\left( 1-x\right) x\left. \frac{\partial \Delta \overline{%
\overline{c}}_{2}\left( r_{12};n\right) }{\partial n}\right\vert _{\overline{%
\rho }_{0}+x\left( \frac{n\left( \mathbf{r}_{1}\right) +n\left( \mathbf{r}%
_{2}\right) }{2}-\overline{\rho }_{0}\right) } \\
&=&\left( \frac{n\left( \mathbf{r}_{1}\right) +n\left( \mathbf{r}_{2}\right) 
}{2}-\overline{\rho }_{0}\right) ^{-1}\int_{0}^{1}dx\left( 1-x\right) x\frac{%
\partial }{\partial x}\Delta c_{2}\left( r_{12};\overline{\rho }_{0}+x\left( 
\frac{n\left( \mathbf{r}_{1}\right) +n\left( \mathbf{r}_{2}\right) }{2}-%
\overline{\rho }_{0}\right) \right)   \notag \\
&=&\left( \frac{n\left( \mathbf{r}_{1}\right) +n\left( \mathbf{r}_{2}\right) 
}{2}-\overline{\rho }_{0}\right) ^{-1}\int_{0}^{1}dx\left( -1+2x\right)
\Delta c_{2}\left( r_{12};\overline{\rho }_{0}+x\left( \frac{n\left( \mathbf{%
r}_{1}\right) +n\left( \mathbf{r}_{2}\right) }{2}-\overline{\rho }%
_{0}\right) \right)   \notag
\end{eqnarray}%
So%
\begin{eqnarray}
&&-\frac{1}{2}\int_{V}\int_{V}\left( n\left( \mathbf{r}_{1}\right) -%
\overline{\rho }_{0}\right) \left( n\left( \mathbf{r}_{2}\right) -n\left( 
\mathbf{r}_{1}\right) \right) \Delta \overline{\overline{c}}_{2}\left(
r_{12};n\left( \mathbf{r}_{1}\right) ,\overline{\rho }_{0}\right) d\mathbf{r}%
_{1}d\mathbf{r}_{2} \\
&=&\frac{1}{4}\int_{V}\int_{V}\left( n\left( \mathbf{r}_{2}\right) -n\left( 
\mathbf{r}_{1}\right) \right) ^{2}\int_{0}^{1}dx\left( 1-x\right) \Delta
c_{2}\left( r_{12};\overline{\rho }_{0}+x\left( \frac{n\left( \mathbf{r}%
_{1}\right) +n\left( \mathbf{r}_{2}\right) }{2}-\overline{\rho }_{0}\right)
\right)   \notag \\
&&-\frac{1}{2}\int_{V}\int_{V}\left( \frac{n\left( \mathbf{r}_{1}\right)
+n\left( \mathbf{r}_{2}\right) }{2}-\overline{\rho }_{0}\right) \left(
n\left( \mathbf{r}_{2}\right) -n\left( \mathbf{r}_{1}\right) \right) \Delta 
\overline{\overline{c}}_{2}\left( r_{12};\frac{n\left( \mathbf{r}_{1}\right)
+n\left( \mathbf{r}_{2}\right) }{2},\overline{\rho }_{0}\right) d\mathbf{r}%
_{1}d\mathbf{r}_{2}  \notag \\
&&+\frac{1}{4}\int_{V}\int_{V}\left( n\left( \mathbf{r}_{2}\right) -n\left( 
\mathbf{r}_{1}\right) \right) ^{2}\int_{0}^{1}dx\left( -1+2x\right) \Delta
c_{2}\left( r_{12};\overline{\rho }_{0}+x\left( \frac{n\left( \mathbf{r}%
_{1}\right) +n\left( \mathbf{r}_{2}\right) }{2}-\overline{\rho }_{0}\right)
\right)   \notag \\
&&+O\left( \frac{n\left( \mathbf{r}_{1}\right) -n\left( \mathbf{r}%
_{2}\right) }{2}\right) ^{3}  \notag
\end{eqnarray}%
The second term on the right is odd under an interchange of the indices and
so vanishes. What is left gives%
\begin{eqnarray}
&&-\frac{1}{2}\int_{V}\int_{V}\left( n\left( \mathbf{r}_{1}\right) -%
\overline{\rho }_{0}\right) \left( n\left( \mathbf{r}_{2}\right) -n\left( 
\mathbf{r}_{1}\right) \right) \Delta \overline{\overline{c}}_{2}\left(
r_{12};n\left( \mathbf{r}_{1}\right) ,\overline{\rho }_{0}\right) d\mathbf{r}%
_{1}d\mathbf{r}_{2} \\
&=&\frac{1}{4}\int_{V}\int_{V}\left( n\left( \mathbf{r}_{2}\right) -n\left( 
\mathbf{r}_{1}\right) \right) ^{2}\int_{0}^{1}x\Delta c_{2}\left( r_{12};%
\overline{\rho }_{0}+x\left( \frac{n\left( \mathbf{r}_{1}\right) +n\left( 
\mathbf{r}_{2}\right) }{2}-\overline{\rho }_{0}\right) \right) dx  \notag \\
&&+O\left( \frac{n\left( \mathbf{r}_{1}\right) -n\left( \mathbf{r}%
_{2}\right) }{2}\right) ^{3}  \notag
\end{eqnarray}%
The expression for the free energy is thus%
\begin{eqnarray}
\beta F_{ex}\left[ n\right]  &=&\beta F_{ex}^{HS}\left[ n\right] +\int d%
\mathbf{r}\;\Delta f\left( n\left( \mathbf{r}\right) \right)  \\
&&+\frac{1}{4}\int_{V}\int_{V}\left( n\left( \mathbf{r}_{1}\right) -n\left( 
\mathbf{r}_{2}\right) \right) ^{2}\Delta c_{2}^{tail}\left( r_{12}\right) d%
\mathbf{r}_{2}d\mathbf{r}_{1}  \notag \\
&&+\frac{1}{4}\int_{V}\int_{V}\left( n\left( \mathbf{r}_{1}\right) -n\left( 
\mathbf{r}_{2}\right) \right) ^{2}\left[ 2\int_{0}^{1}\lambda \Delta
c_{2}^{core}\left( r_{12};\overline{\rho }_{0}+\lambda \left( \frac{n\left( 
\mathbf{r}_{1}\right) +n\left( \mathbf{r}_{2}\right) }{2}-\overline{\rho }%
_{0}\right) \right) d\lambda \right] d\mathbf{r}_{1}d\mathbf{r}_{2}  \notag
\\
&&+O\left( \frac{n\left( \mathbf{r}_{1}\right) -n\left( \mathbf{r}%
_{2}\right) }{2}\right) ^{3}  \notag
\end{eqnarray}%
or%
\begin{eqnarray}
\beta F_{ex}\left[ n\right]  &=&\beta F_{ex}^{HS}\left[ n\right] +\int d%
\mathbf{r}\;\Delta f\left( n\left( \mathbf{r}\right) \right)  \\
&&+\frac{1}{4}\int_{V}\int_{V}\left( n\left( \mathbf{r}_{1}\right) -n\left( 
\mathbf{r}_{2}\right) \right) ^{2}\left[ 2\int_{0}^{1}\lambda \Delta
c_{2}\left( r_{12};\overline{\rho }_{0}+\lambda \left( \frac{n\left( \mathbf{%
r}_{1}\right) +n\left( \mathbf{r}_{2}\right) }{2}-\overline{\rho }%
_{0}\right) \right) d\lambda \right] d\mathbf{r}_{1}d\mathbf{r}_{2}  \notag
\\
&&+O\left( \frac{n\left( \mathbf{r}_{1}\right) -n\left( \mathbf{r}%
_{2}\right) }{2}\right) ^{3}  \notag
\end{eqnarray}

\bigskip

\bigskip 
\bibliographystyle{apsrev}
\bibliography{../../physics}

\bigskip

\section{Figure captions}

Fig.1. The coexistence curve for the Lennard-Jones fluid as calculated
using both the WCA perturbation theory and the BH theory. The full lines are
the liquid-vapor coexistence curves, the dashed-lines are the spinodals and
the symbols are the simulation data from ref.\protect\cite%
{HansenLJPhaseDiagram}(circles) and from ref. \protect\cite%
{Potoff_LJ_Interface} (squares).

Fig. 2. The DCF at zero density and $T^{\ast }=1.5$. The symbols are the the negative of the Meyer function (the exact
result for zero density), and the lines are from the
three choices of tail function described in the text. Figure (a) is based on
the exact equation of state while in Fig.(b), the Barker-Henderson
perturbation theory is used.

Fig. 3. The DCF at $T^{\ast }=1.5$ and $\protect\rho \protect\sigma %
^{3}=0.4 $ as determined from the model using the Barker-Henderson
perturbation theory, lines, and the simulation data of Llano-Restrepo and
Chapman\protect\cite{DCF-Simulation}.

Fig. 4. The DCF at $T^{\ast }=1.5$ and $\protect\rho \protect\sigma %
^{3}=0.6 $ as determined from the model using the Barker-Henderson
perturbation theory, lines, and the simulation data of Llano-Restrepo and
Chapman\protect\cite{DCF-Simulation}.

Fig. 5. The DCF at $T^{\ast }=1.5$ and $\protect\rho \protect\sigma %
^{3}=0.9 $ as determined from the model using the Barker-Henderson
perturbation theory, lines, and the simulation data of Llano-Restrepo and
Chapman\protect\cite{DCF-Simulation}. The hard-sphere contribution to
the DCF is shown in black.

Fig. 6. The DCF at $T^{\ast }=0.72$ and $\protect\rho \protect\sigma %
^{3}=0.85$ as determined from the model using the Barker-Henderson
perturbation theory, lines, and the simulation data of Llano-Restrepo and
Chapman\protect\cite{DCF-Simulation}.The hard-sphere contribution to
the DCF is shown in black.

Fig. 7. The surface tension as a function of temperature. The symbols are
measurements from simulations (circles from ref.\protect\cite%
{Duque-LJ-interface},squares from ref.\protect\cite{Mecke-LJ_Interface} and
triangles from ref. \protect\cite{Potoff_LJ_Interface}). The lines are from
the LDCF-II DFT evaluated with the corrected mean-field DCF using the
equation of state calculated from the BH perturbation theory (full line) and
the WCA theory (broken line).

Fig. 8. The density profiles calculated from the various DFTs for $%
T^{\ast}=0.7$.

Fig. 9. A comparison of the surface tension as a function of reduced
temperature as calculated from the LDCF-II, ESS, Tang and SGA DFTs.

Fig. 10. The variation of the reduced surface tension, $\protect\gamma %
^{\ast } $ as a function of the reference density $\protect\rho _{0}\protect%
\sigma ^{3}$ for the different DFTs using the BH equation of state and for $%
T^{\ast}=0.7 $. The vertical lines are the boundary of the region for which
a stable solution was found.

\newpage

\begin{figure*}[h!tb] \centering
\resizebox{12cm}{!}{\centerline{\includegraphics[angle=-90]{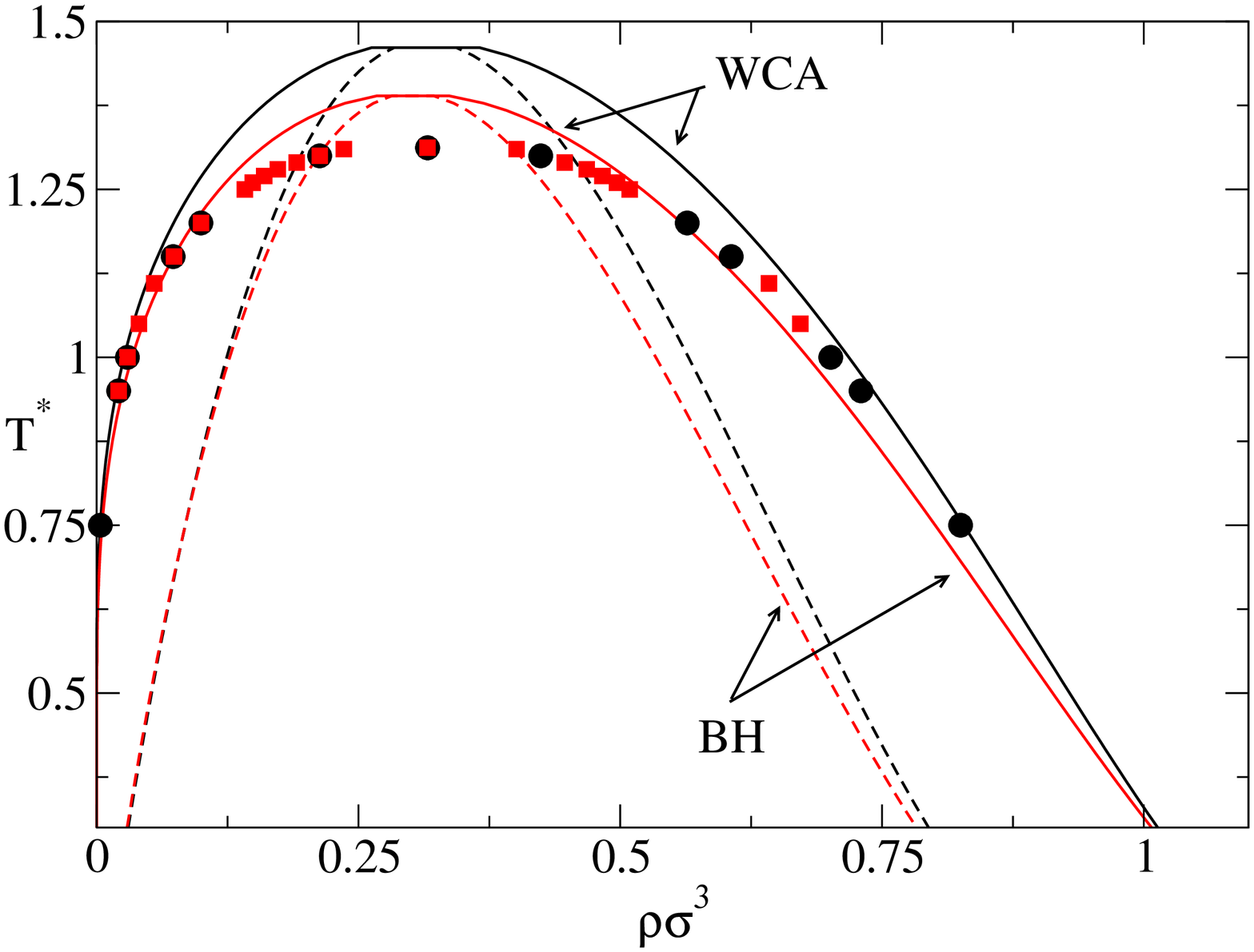}}}
\caption{}
\label{fig1}
\end{figure*}

\begin{figure*}[h!tb] \centering
\resizebox{12cm}{!}{\centerline{\includegraphics[angle=-90]{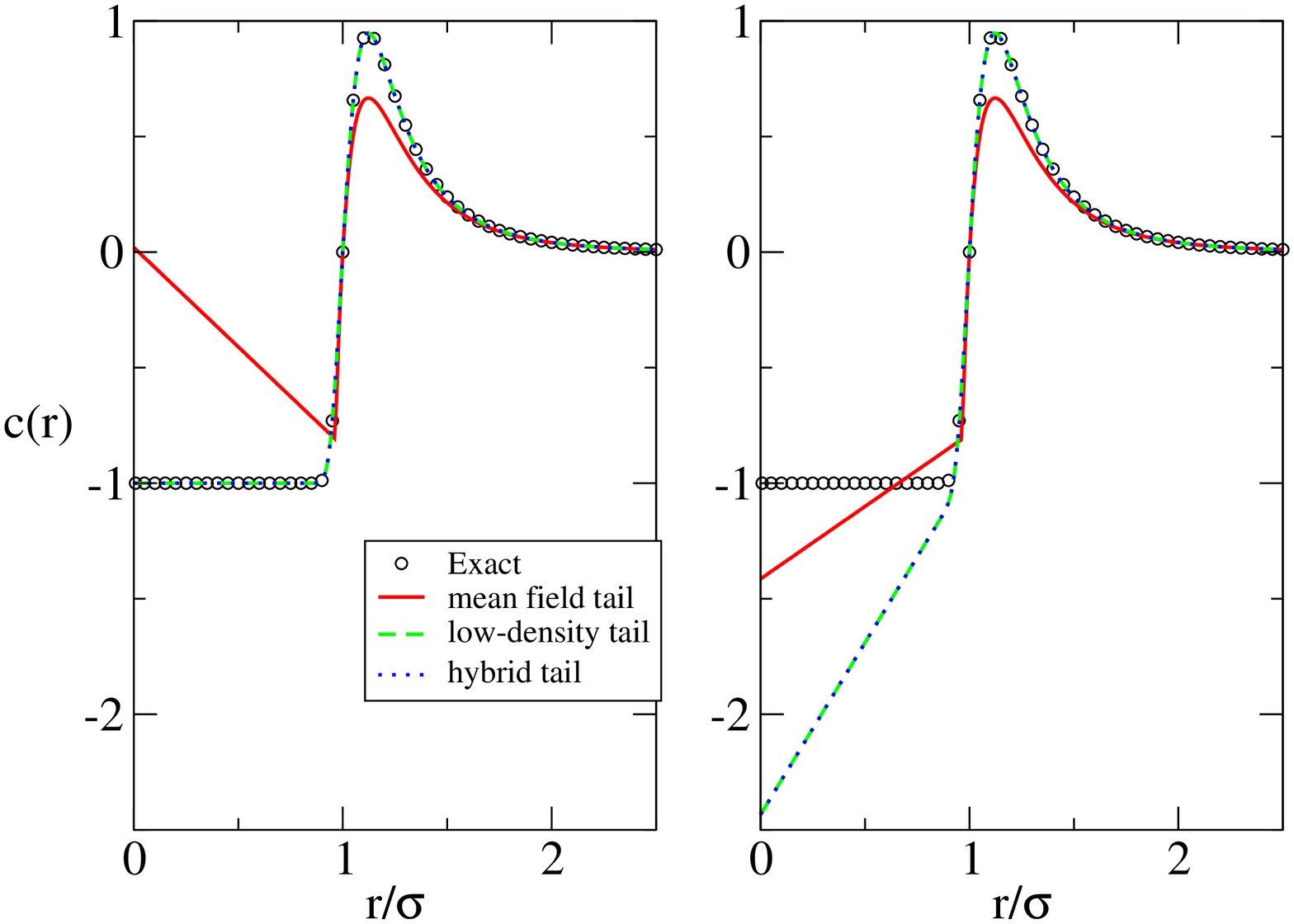}}}
\caption{}
\label{fig2}
\end{figure*}

\begin{figure*}[h!tb] \centering
\resizebox{12cm}{!}{\centerline{\includegraphics[angle=-90]{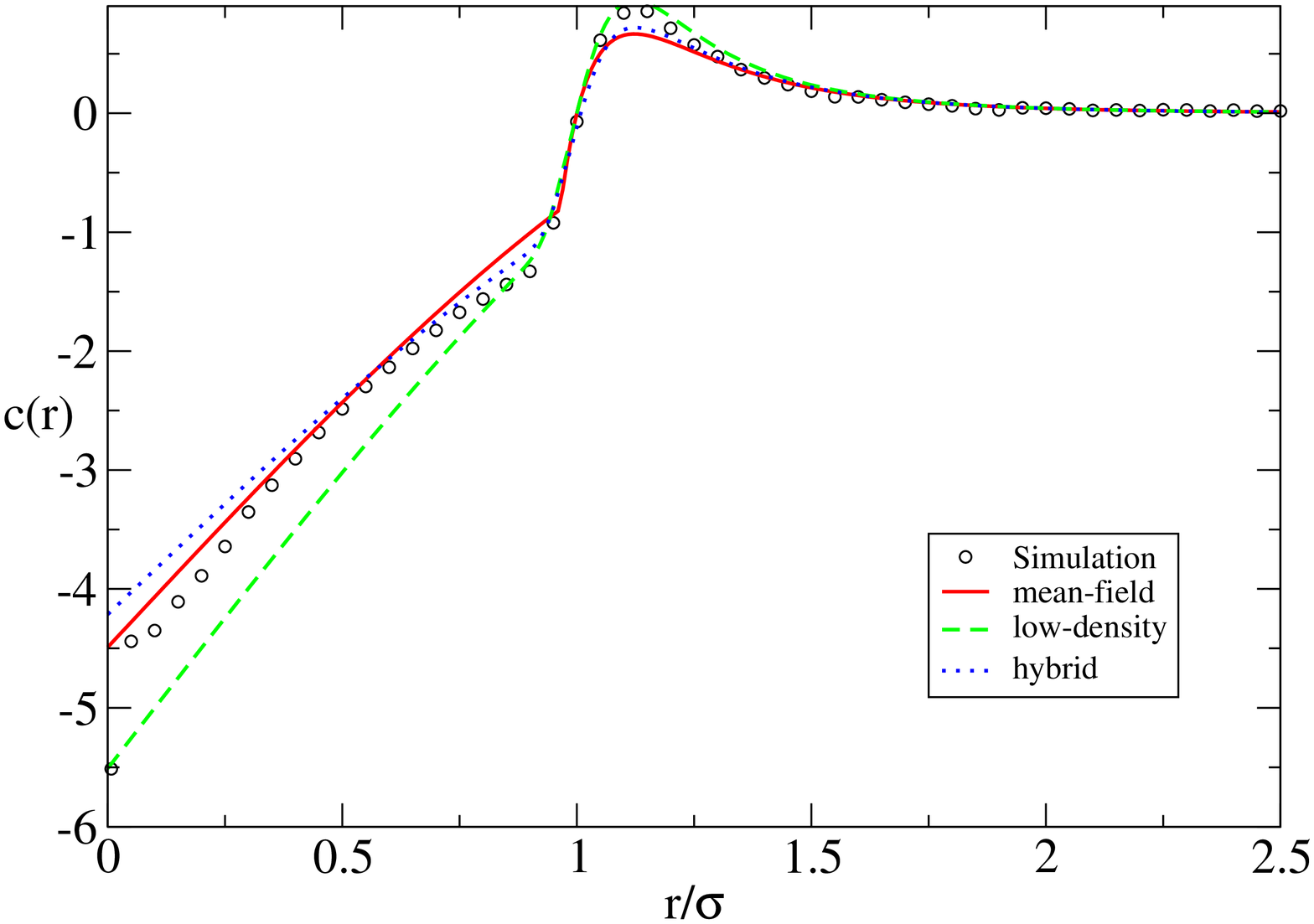}}}
\caption{}
\label{fig3}
\end{figure*}

\begin{figure*}[h!tb] \centering
\resizebox{12cm}{!}{\centerline{\includegraphics[angle=-90]{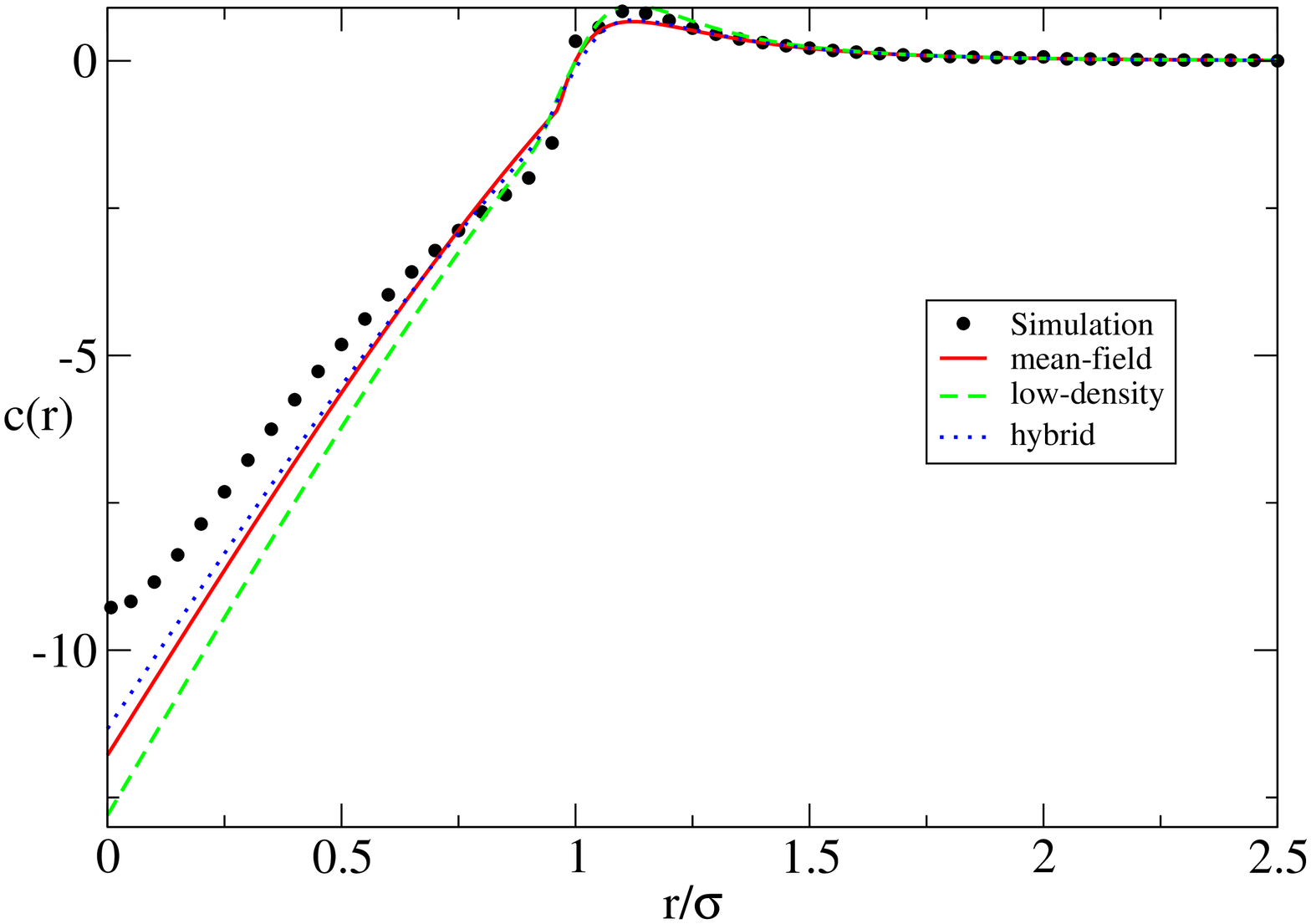}}}
\caption{}
\label{fig4}
\end{figure*}

\begin{figure*}[h!tb] \centering
\resizebox{12cm}{!}{\centerline{\includegraphics[angle=-90]{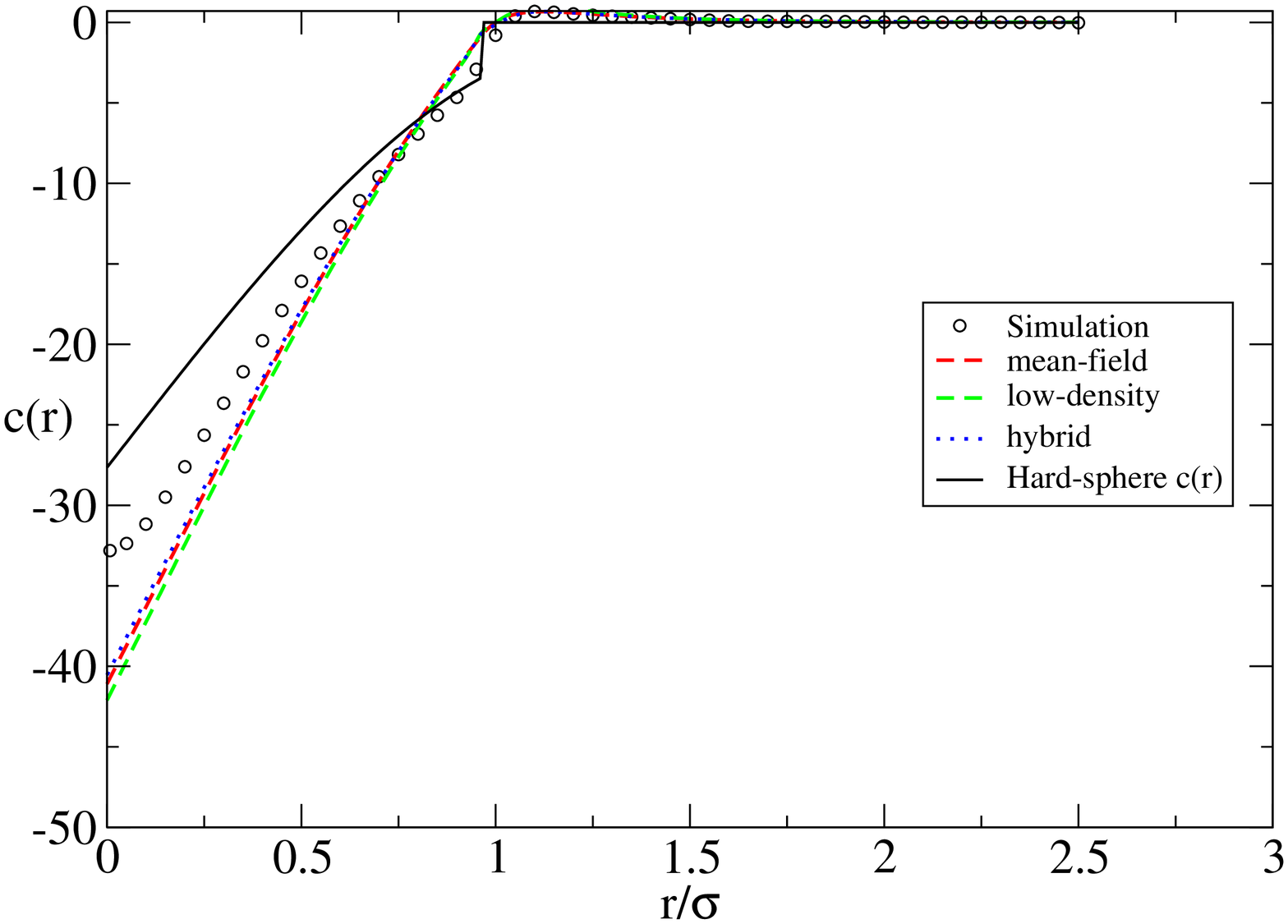}}}
\caption{}
\label{fig5}
\end{figure*}

\begin{figure*}[h!tb] \centering
\resizebox{12cm}{!}{\centerline{\includegraphics[angle=-90]{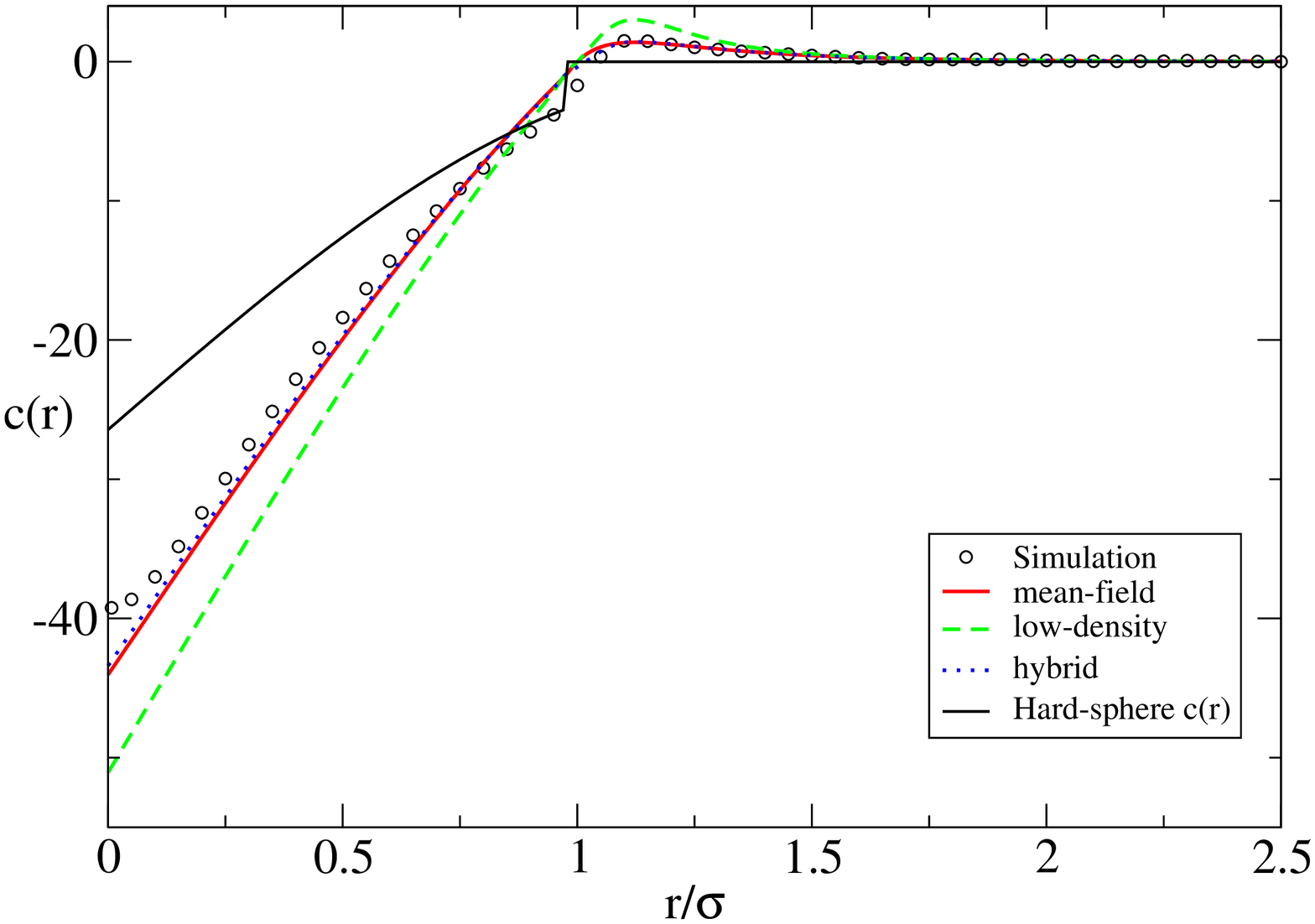}}}
\caption{}
\label{fig6}
\end{figure*}

\begin{figure*}[h!tb] \centering
\resizebox{12cm}{!}{\centerline{\includegraphics[angle=-90]{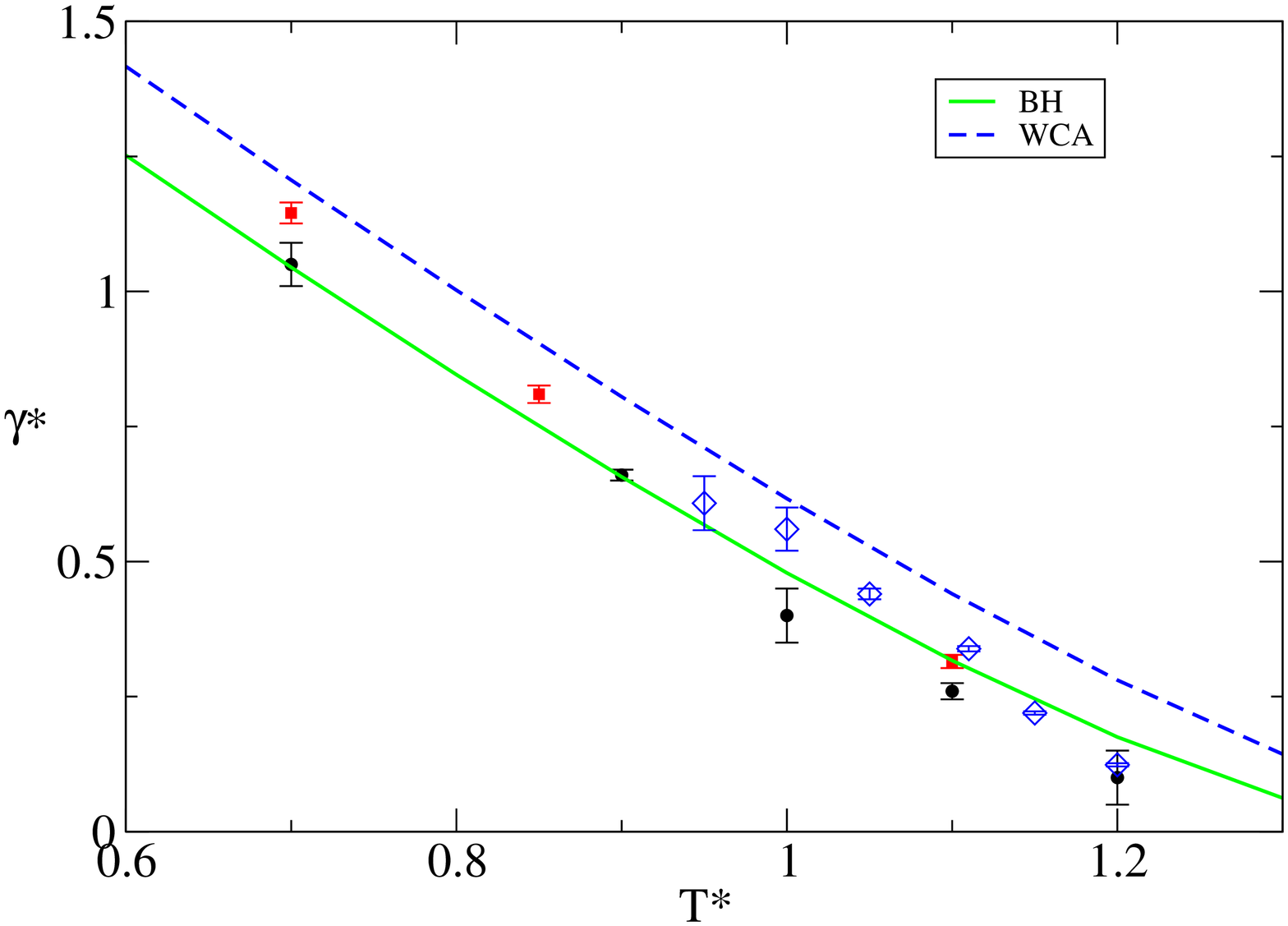}}}
\caption{}
\label{fig7}
\end{figure*}

\begin{figure*}[h!tb] \centering
\resizebox{12cm}{!}{\centerline{\includegraphics[angle=-90]{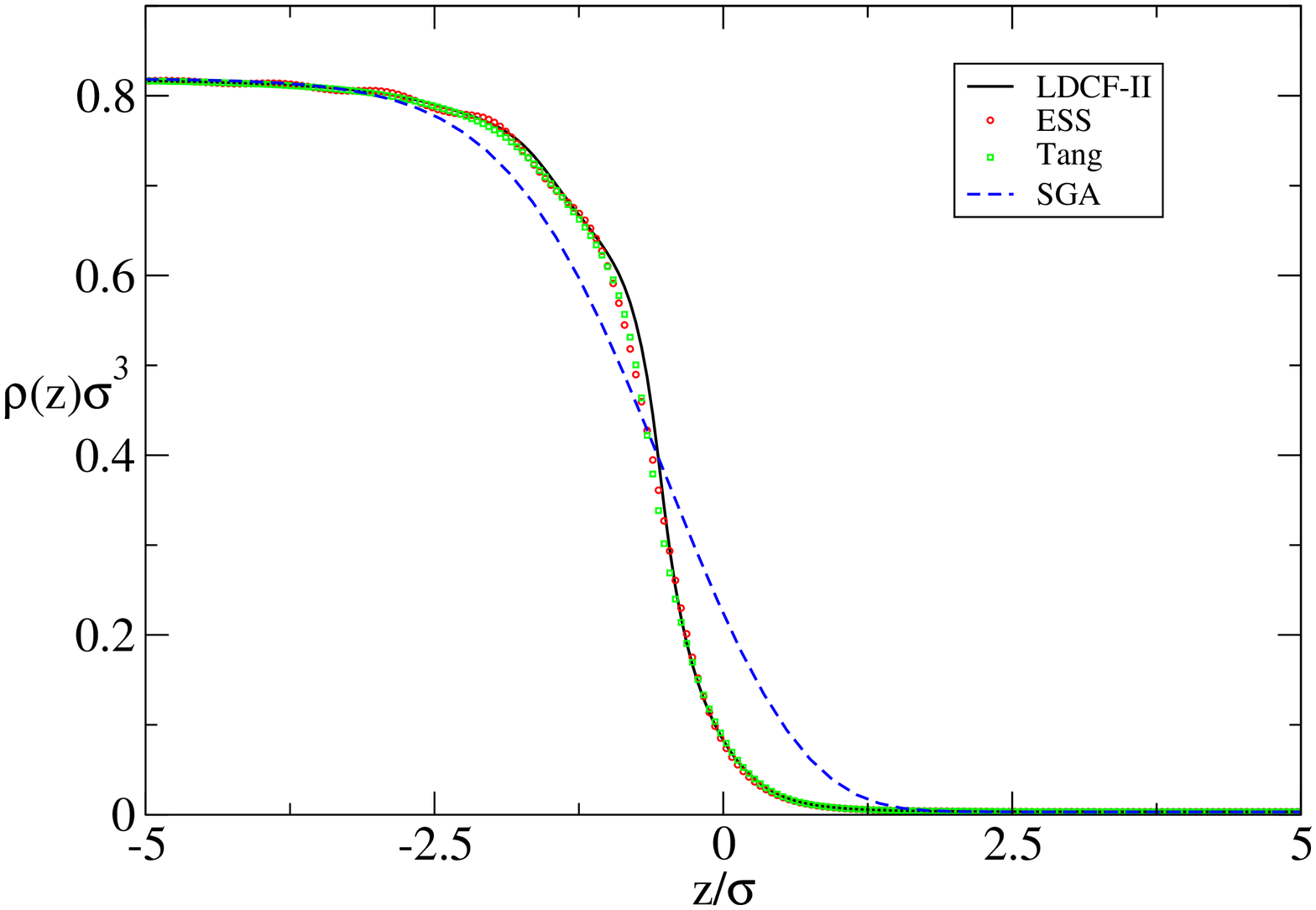}}}
\caption{}
\label{fig8}
\end{figure*}

\begin{figure*}[h!tb] \centering
\resizebox{12cm}{!}{\centerline{\includegraphics[angle=-90]{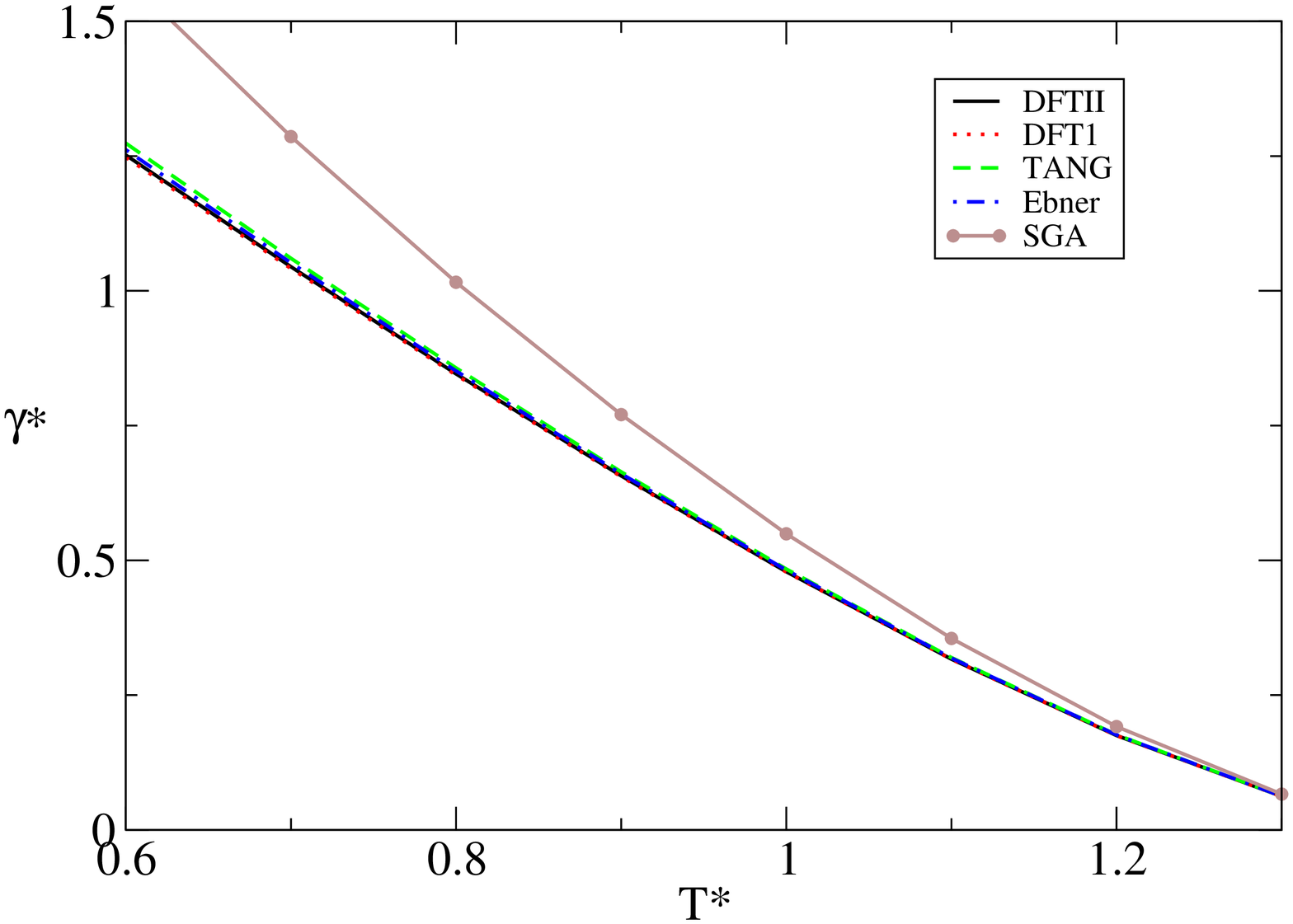}}}
\caption{}
\label{fig9}
\end{figure*}

\begin{figure*}[h!tb] \centering
\resizebox{12cm}{!}{\centerline{\includegraphics[angle=-90]{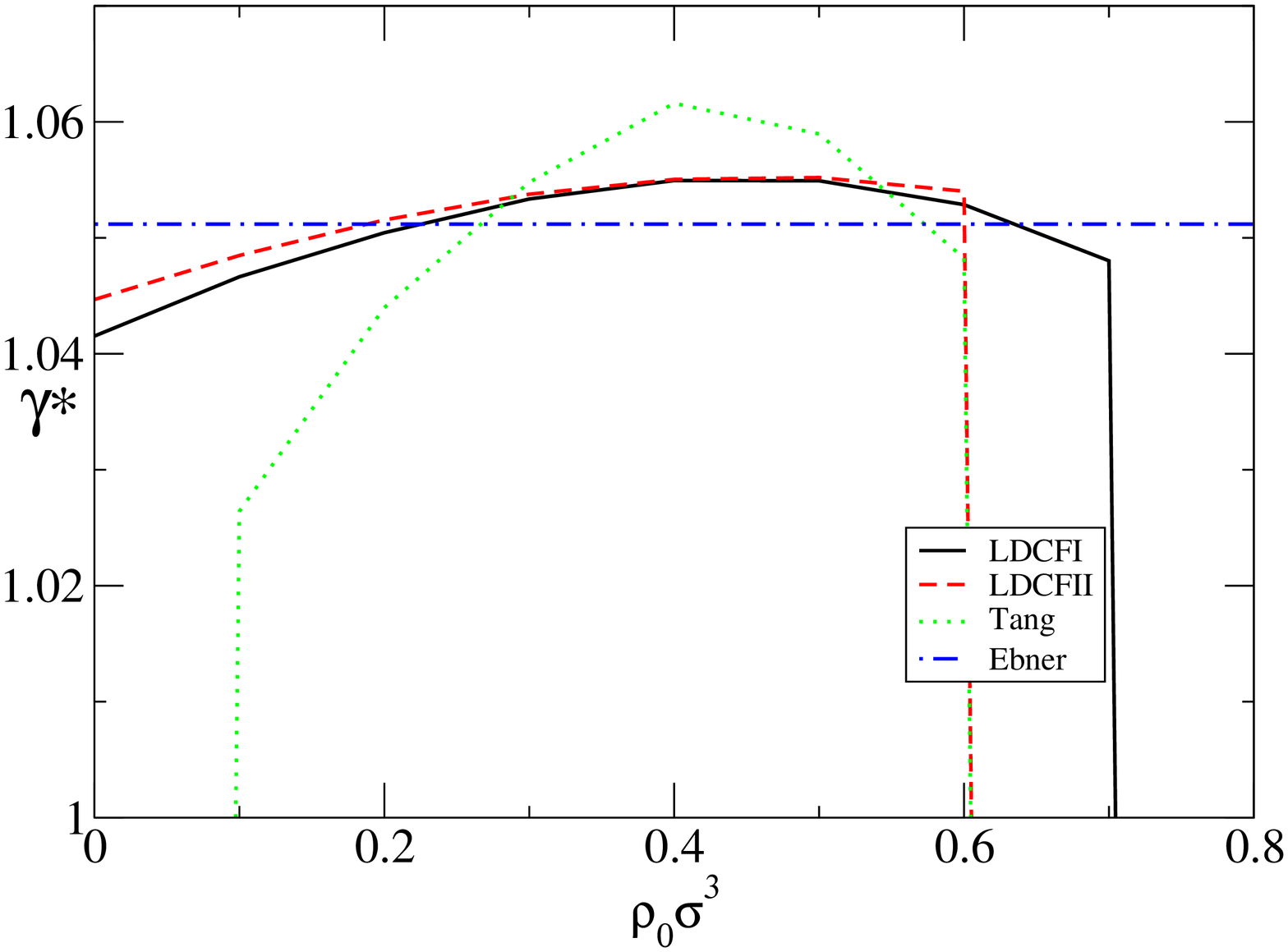}}}
\caption{}
\label{fig9}
\end{figure*}

\end{document}